\documentclass[aps,pra,showpacs,onecolumn, amsfonts,amssymb,amsmath,preprint]{revtex4}  

\usepackage{graphicx} 
\usepackage{array}
\usepackage{color}
\usepackage{float}
\usepackage{subfigure}
\usepackage{color}
\usepackage{multirow}

\newcommand{\ketO}[0]{$\left|0\right>$\,}
\newcommand{\ketI}[0]{$\left|1\right>$\,}
\newcommand{\ketE}[0]{$\left|\mathrm{e}\right>$\,}

\begin{document}

\title{Robust pulses for high fidelity non-adiabatic geometric gate operations in an off-resonant three-level system} 

\author{Ying Yan$^{1,}$$^2$, Jie Lu$^{3,}$$^4$, Lin Wan$^{1,}$$^2$, Joel Moser$^{1,}$$^2$} %
\affiliation{$^1$ School of Optoelectronic Science and Engineering \& Collaborative Innovation Center of Suzhou Nano Science and Technology, Soochow University, Suzhou 215006, China.}
\affiliation{$^2$ Key Lab of Advanced Optical Manufacturing Technologies of Jiangsu Province \& Key Lab of Modern Optical Technologies of Education Ministry of China, Soochow University, 215006 Suzhou, China}
\affiliation{$^3$ Department of Physics, Shanghai University, 200444 Shanghai, China}
\affiliation{$^4$ Shanghai Key Lab for Astrophysics, 100 Guilin Road, 200234 Shanghai, China}

\renewcommand{\thefootnote}{\fnsymbol{footnote}}
\footnotetext[1]{yingyan@suda.edu.cn, moser.joel@gmail.com}

\date{\today}

\begin{abstract}
  We propose a method to design pulses in a resonant three-level system to enhance the robustness of non-adiabatic geometric gate operations. By optimizing the shape of the pulse envelope, we show that the gate operations are more robust against frequency detuning than they are with Gaussian and square pulses. Our method provides a way to design pulses that can be employed in a system where robustness against frequency variations or inhomogeneous broadening is required, and may be extended to ensure robustness against other physical imperfections such as intensity fluctuations and random noises.\\

\end{abstract}

\maketitle

\section{Introduction}

Controlling the state of qubits precisely is essential for efficient quantum computing. In systems that suffer from physical imperfections, such as variations in qubit addressing frequencies and detuning due to inhomogeneous broadening, high fidelity gate operations remain challenging. For example, in superconducting transmon qubits, the qubit parameters (e.g. the qubit frequency) may drift or fluctuate over time \cite{Devoret2004}. In addition, the ancilla qubits that are all built in the same circuit for the purpose of performing error correction are never perfectly identical. For an accurate control, the addressing frequency of each ancilla qubit has to be measured with high precision, and control signals have to be tailored to match individual qubits \cite{Devoret2013, Nicolas2018}. Another example is the system of qubits represented by an ensemble of rare-earth-ions having an inhomogeneous broadening of about 170 kHz \cite{Lars2008}. Using robust pulses that are insensitive to variations in the system allows for substantial improvement, as we show below.  

Holonomic quantum computing is one of the approaches for constructing robust and fast quantum gate operations based on geometric phases \cite{Zanardi1999}. It has received increasing attention for its intrinsic robustness against errors accrued as the quantum system evolves. However, the adiabatic feature in the original scheme limits its practical applications because of the relatively long operation time compared to the coherence time of the system \cite{Duan2001,Faoro2003, Solinas2003}. After the scheme was generalized to non-adiabatic, non-Abelian geometric gates \cite{Erik2012}, experimental demonstration of holonomic quantum gates was reported in various physical systems, including three-level superconducting artificial atoms \cite{Abdumalikov2013} and electron spins of nitrogen-vacancy centers \cite{CZu2014,Silvia2014}. Two conditions have to be met to realize non-adiabatic geometric gates \cite{Erik2012}. Firstly, the parallel-transport condition has to hold at any time during the state evolution, that is, the dynamical phase has to be zero at any time. Secondly, the integral of the time-dependent part of the Rabi frequency over the operation time has to be equal to $\pi$. In a resonant three-level system, envelope of pulses can have an arbitrary shape as long as the above conditions are met \cite{Emmi2016}. However, for an off-resonant system where the driving frequency is detuned from the qubit transition frequency due to inhomogeneous broadening or fluctuations, a square pulse has to be used in order to maintain the geometric nature of the evolution, which requires the system Hamiltonian to commute with the time evolution operator \cite{Erik2016b,GFXu2015}. Yet, the use of square pulses has several disadvantages: (i) it makes it impossible to utilize the pulse envelope to optimize the pulse performance against variations of some physical quantities from a perfect situation \cite{Lars2008,Roos2004}; (ii) the Fourier transform in frequency domain of a square pulse has multiple frequency components, which might excite unwanted transitions; (iii) gate operations using square pulses are sensitive to fluctuations in intensity and phases. Using a hyperbolic secant function as a pulse shape, arbitrary single qubit gate operations have been demonstrated in an off-resonant system \cite{Lars2008}. However, the gate operates slowly as the pulses are adiabatic.

Here we propose an alternative type of pulses constructed in a resonant three-level system in which the pulse envelope can be tailored through multiple degrees of freedom. Pulses with optimized envelope can perform high fidelity non-adiabatic geometric gate operations in a slightly off-resonant system, both in the presence of a detuning between laser frequency and the qubit transition and in the presence of fluctuations of the qubit addressing frequency. Fidelity of 99\% over a frequency detuning range of $\pm$410 kHz is accomplished with two pairs of pulses implemented consecutively. One performs on the bright state and leave the dark state untouched, and the other works on the initial dark state and leaves the initial bright state untouched. As a result, the frequency-detuning dependent phase evolution factor accumulates equally in both states and can be seen as a global phase \cite{Roos2004,Lars2008}. Our finding provides an alternative way of designing pulses for performing high-fidelity geometric gate operations in systems where the qubit is addressed in frequency, and may be extended to enforce robustness against other physical parameters such as intensity fluctuations and random noises. 

The article is organized as follows. Section II describes the model of pulses we propose for $\sigma_x$, $\sigma_y$, $\sigma_z$ and Hadamard gate operations. Section III is devoted to introducing the performance of the pulses in simulations for a slightly off-resonant system. In Section IV we discuss our results and present our conclusion.

\section{Light Pulses for Geometric Gate operations}

The light pulses we propose are constructed from an on-resonant three-level system depicted in Figure 1, where the qubit is represented by two ground state levels $\left|0\right>$ and $\left|1\right>$. These two levels are coupled through optical light fields $\Omega_0 = 2A\Omega(t)e^{-i\varphi}$ and $\Omega_1 = 2B\Omega(t)$ via the excited state $\left|\rm{e}\right>$, where $A = \sin(\theta/2)$ and $B = -\cos(\theta/2)$ are two constants, and $\varphi$ denotes the time-independent phase difference between the two fields. $\Omega(t)$ is the real-valued Rabi frequency describing the interaction between the light fields and the quantum system, defined as $\Omega(t) = -\vec{\mu}\cdot\vec{E}/\hbar$, where $\vec{\mu}$ and $\vec{E}$ denote the transition dipole moment and the electric field, respectively. In the rotating wave approximation, the Hamiltonian of the system in the basis formed by \ketI, \ketE and \ketO reads

\begin{equation}
\begin{aligned}
H(t) & = \frac{\hbar}{2}\begin{bmatrix}  
0 & \Omega_1(t) & 0 \\
\Omega_1(t) & 0 & \Omega_0^*(t)\\
0 & \Omega_0(t) & 0
\end{bmatrix},\\
\end{aligned}
\end{equation}
where $\Omega_0^*(t)$ is the complex conjugate of $\Omega_0(t)$.

\begin{figure}[H]
\centering
\begin{minipage}{5cm}
\centering
	\includegraphics[width=3.5cm,height=4.5cm]{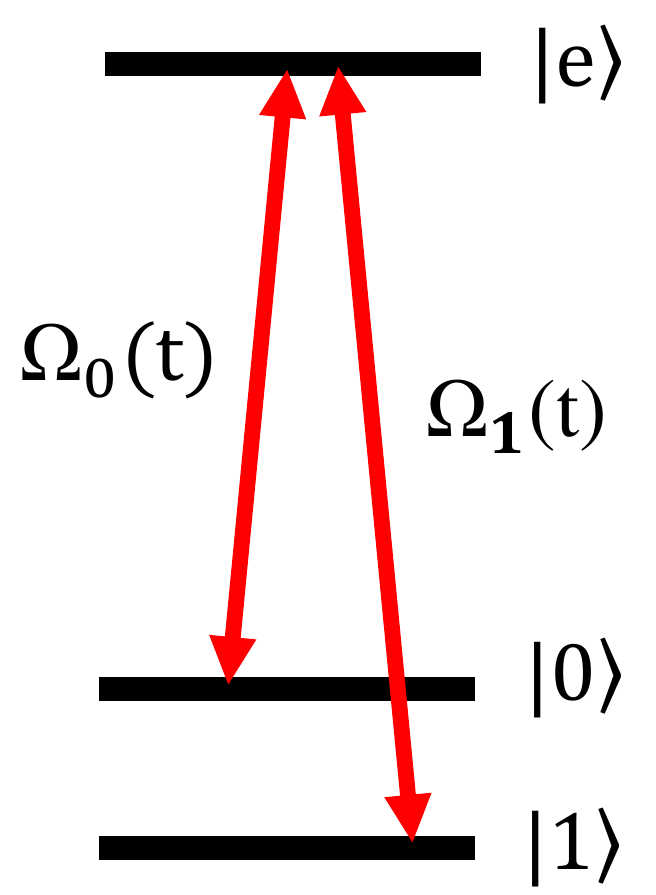}
\end{minipage}
	\caption{(Color online) A shematic 3-level system. The qubit is represented by ground state levels $\left|0\right>$ and $\left|1\right>$, which are coupled through optical transitions $\left|0\right>\rightarrow\left|\rm{e}\right>$ and $\left|1\right>\rightarrow\left|\rm{e}\right>$. $\Omega_1$ and $\Omega_0$ are the respective Rabi frequencies. $\varphi$ is a time-independent phase factor.}
	\label{fig.1}
\end{figure}

As the Hamiltonian $H(t)$ shown in Eq. (1) commutes at any two different times, the time evolution operator of the three-level system can be written as $U(t_1,0) = e^{-\frac{i}{\hbar}\int^{t_1}_0 H(t)dt} = \left|d\right>\left<d\right| - \left|b\right>\left<b\right|$ under the conditions that $\alpha(t_1) = \int^{t_1}_0 \Omega(t) dt = \pi$ \cite{Erik2012}, where $t_1$ is the pulse duration, and $\left|d\right> = -B \left|0\right> + A e^{i\varphi} \left|1\right>$ and $\left|b\right> = A e^{-i\varphi} \left|0\right> + B \left|1\right>$ are the dark and bright states, respectively. It can also be shown that the dynamical phase $\gamma_d(t) = -\frac{1}{\hbar}\int_0^{t} \left<k(t)|H(t)|l(t)\right> dt \equiv 0$ ($k, l = b, d$) with $t\in [0,t_1]$. This means that the time evolution of any qubit state in a three-level system is a pure geometric evolution, and the state returns to the qubit subspace in the end. Such an evolution in qubit subspace is described by $U^{(1)} = P_c\cdot U(t_1,0)\cdot P_c = \hat{n}\cdot\vec{\sigma} = i e^{-i\pi\hat{n}\cdot\vec{\frac{\sigma}{2}}}$ where $P_c = \left|0\right>\left<0\right| + \left|1\right>\left<1\right|$, $\hat{n}= (\sin \theta\cos\varphi, \sin\theta\sin\varphi, \cos\theta)$, and $\vec{\sigma} = (\sigma_x, \sigma_y, \sigma_z)$ with $\sigma_i$ ($i = x, y, z$) is a Pauli matrix. It is a rotation of the qubit state about a unitary vector $\hat{n}$ by an angle of $\pi$. Depending on the specific values of $\theta$ and $\varphi$, the time evolution operator $U^{(1)}$ performs various single-qubit gate operations as shown in Table I.

\begin{table} [htbp] 
  \centering
  \caption{Values of pulse parameters $\theta$ and $\varphi$ for $\sigma_x$, $\sigma_y$, $\sigma_z$ and Hadamard gate operators. }\label{Table 1}}
  \centering
  \setlength{\tabcolsep}{8mm}{
  \scalebox{1}{
  \begin{tabular}{c|c c } 
  \hline
  \hline
  Operator & $\theta$ & $\varphi$\\
  \hline
  $\sigma_x$ & $\pi$/2 & 0 \\
  \hline
  $\sigma_y$ & $\pi$/2 & $\pi$/2 \\
  \hline
  $\sigma_z$ & 0 & arbitrary \\
  \hline
  Hadamard  & $\pi$/4 & 0 \\
   \hline
   \hline
  \end{tabular}}
\end{table}

For an off-resonant three-level system, the time evolution operator $U(t_1,0)$ mentioned above is incorrect as the Hamiltonian at different times does not commute, unless in the case of a square pulse \cite{Erik2016b}. However, if the pulse envelope is a summation of multiple Cosine components as in the following expression
\begin{equation}
\Omega(t) = \frac{\pi}{t_1} + \sum^\infty_{n = 1} a_n\cdot \frac{n\pi}{t_1}\cdot\cos(\frac{n\pi}{t_1}t),
\end{equation}
where the parameters $a_n$ are to be determined, pulses developed in a resonant system as shown in Figure 1 can be robust against frequency detuning over a certain frequency range. Such pulses can be applied to slightly off-resonant systems. A form similar to Eq. (2) was used to develop robust pulses for NOT gate \cite{Leo2017}, but there the gate was not geometric. Such an envelope meets the criterion $\int^{t_1}_0\Omega(t)dt= \pi$. For an adiabatic geometric gate operation, the Hamiltonian needs to return to its original value so that the state undergoes a cyclic evolution \cite{Aharronov1987}, whereas non-adiabatic geometric gate does not require this condition. However, for the sake of a successful implementation of the pulses in experiments, one would prefer $\Omega(0) = \Omega(t_1) = 0$ (where $t_1$ is the pulse duration) as sudden rises of pulses in time domain could have multiple components in frequency domain, which may excite unwanted transitions. This condition requires that
\begin{equation}
\begin{aligned}
& a_1 + 3a_3+5a_5 + 7a_7 + ...+ (2k-1)\cdot a_{2k-1} = 0\\
& a_2 + 2a_4 + 3a_6 + 4a_8 + ... + (2k)\cdot a_{2k} = -0.5,
\end{aligned}
\end{equation}
where $k$ is a positive integer that can be infinitely large in principle, but might in practice be limited by the rise time of the instrument that generates the pulses. Here we consider a maximum value of $k = 4$. The 6 degrees of freedom available in Eq. (3) can be used to tailor the pulse envelope and achieve the desired robustness.

However, if one were to implement above pulses with their envelopes optimized, robustness could still not be achieved because these pulses would only interact with $\left|b\right>$ state, but not with $\left|d\right>$. A detuning-dependent phase factor would be accumulated on $\left|b\right>$ but not on $\left|d\right>$. Therefore, another pair of phase-compensation pulses has to be implemented to only interact with $\left|d\right>$, and let it acquire the same phase as that of $\left|b\right>$ during the gate operation time. The phase can then be seen as a global phase \cite{Roos2004}. Explicitly, the compensation pulse has to meet two requirements. (i) It has to have appropriate $\theta'$ and $\varphi'$ parameters to transform the previous $\left|d\right>$ into a new bright state $\left|b'\right>$, and $\left|b\right>$ into a new dark state $\left|d'\right>$. (ii) It has to perform an identity operation on $\left|b'\right>$, i.e. $U^{(1)}(t_2,t_1)\left|b'\right> = \left|b'\right> $, where $t_2-t_1$ defines the duration time of the compensation pulses. To meet requirement (i), the pulse parameters ($\theta', \varphi'$) of the compensation pulses need to be
\begin{equation}
\begin{aligned}
\theta' & = \pi-\theta\\
\varphi' & = \pi + \varphi. \\
\end{aligned}
\end{equation}
To meet requirement (ii) $U^{(1)}(t_2,t_1) = I$, the pulse envelope $\Omega'(t)$ has to satisfy $\int^{t_2}_{t_1} \Omega'(t) dt = 2\pi$. Assuming that $\Omega'(t)$ reads\\
\begin{equation}
\Omega'(t) = \frac{2\pi}{(t_2-t_1)} + \sum^\infty_{n = 1} a'_n\cdot \frac{n\pi}{t_2-t_1}\cdot\cos(\frac{n\pi}{t_2-t_1}t),
\end{equation}
then $a'_n$ must satisfy the following equalities
\begin{equation}
\begin{aligned}
& a'_1 + 3a'_3+5a'_5 + 7a'_7 + ...+ (2k-1)\cdot a'_{2k-1} = 0\\
& a'_2 + 2a'_4 + 3a'_6 + 4a'_8 + ... + (2k)\cdot a'_{2k} = -1.
\end{aligned}
\end{equation}
It is clear that one solution for $\Omega'(t)$ is $\Omega'(t) = 2\Omega(t)$, but it might not be the only one. Rabi frequencies and phases of the two pairs of pulses are summarized in Table II. In the next section, we present the performance of these pulses in simulation. 

\begin{table} [htbp] 
  \centering
  \caption{Rabi frequencies and phases of the two pairs of two-color pulses. $\Omega_{1}$ ($\Omega_{0}$) denotes the Rabi frequencies for transition $\left|1\right>\rightarrow\left|\rm{e}\right>$ ($\left|0\right>\rightarrow\left|\rm{e}\right>$). $A = \sin\frac{\theta}{2}$, $B = -\cos\frac{\theta}{2}$, $A' = \sin\frac{\theta'}{2}$ , $B' = -\cos\frac{\theta'}{2}$, and $\Omega'(t) = 2\Omega(t)$.}}
  \label{Table 2} 
  \centering
  \setlength{\tabcolsep}{6mm}{
  \scalebox{1}{
  \begin{tabular}{c c c} 
  \hline
  \hline
   & Pulse Pair 1 & Pulse Pair 2\\
  \hline
   $\Omega_{1}$  & $2B\cdot\Omega(t)$ & $2B'\cdot\Omega'(t)$ \\
   $\Omega_{0}$ & $2A\cdot\Omega(t)e^{-i\varphi}$ & $2A'\cdot\Omega'(t)e^{-i\varphi'}$ \\
  \hline
  \hline
  \end{tabular} }
\end{table}

\section{Operational fidelity in a slightly off-resonant system}

Considering an initial qubit state
$\left|\psi(0)\right> =  \cos\theta_0 \left|0\right> + \sin\theta_0 e^{i\varphi_0}\left|1\right>$,
where $\theta_0$ and $\varphi_0$ have arbitrary values in [0, 2$\pi$], the gate operations listed in Table I can be achieved by consecutively implementing the two pairs of two-color pulses shown in Table II. The first pair performs gate operations, and the second pair executes phase compensation on the previous dark state. 

To achieve high robustness in fidelity, the parameters $a_n$ in Eq. (3) have to be optimized in the simulation. This is done with the multi-objective goal attainment function in Matlab by setting the goal of operational fidelity at different detuning frequencies to 1. Fidelity is defined as
\begin{equation}
F = \mid\left<\psi(t_2)|\psi_{\rm{tg}}\right>\mid^2,
\end{equation}
where $\left|\psi_{\rm{tg}}\right> = U^{(1)}(t_2,t_1)U^{(1)}(t_1,0)\left|\psi(0)\right>$ is the target state, $U^{(1)}(t_1,0)$ and $U^{(1)}(t_2,t_1)$ denotes the qubit gate operator listed in Table I and the operator implemented by the phase-compensation pulses, respectively. $\left|\psi(t_2)\right> = [C_{1}(t_2), C_{0}(t_2),C_{\rm{e}}(t_2), ]^T$ refers to the wave function at time $t_2$ when the phase compensation pulses end, \textit{T} denotes transpose of the matrix, and $C_{\rm{n}} (t_2)$ (n = 1, 0, or $\rm{e}$) is the probability amplitude at $t= t_2$. They are numerical solutions to the following three-level coupled differential equations, which resemble the Schr$\rm{\ddot{o}}$dinger equation:
\begin{equation}
\begin{array}{rcl}
\dot{C}_1 &=& -\frac{i}{2}\Omega_1\cdot C_{\rm{e}} \\
\dot{C}_0 &=& -\frac{i}{2}\Omega_0\cdot C_{\rm{e}} \\
\dot{C}_{\rm{e}} &=& - \frac{i}{2}\Omega_1\cdot C_1 - \frac{i}{2}\Omega^*_0\cdot C_0 -i\Delta C_{\rm{e}} 
\end{array}
\end{equation}
where $\Delta$ represents the detuning between the center frequency of the pulses and the transition frequencies between the qubit levels to the excited state.

In this work, we take $\left|\psi(0)\right> = \left|1\right>$ ($\theta_0 = \pi/2$, $\varphi_0 = 0$) as an example. The optimized values of the $a_n$ parameters are shown as Op1 in Table III, considering $t_1 = 4$ $\mu$s. Even smaller values of $t_1$ can be used as long as the maximal instantaneous Rabi frequency and its rise time are realistic. Envelopes of Rabi frequencies, $\Omega_{1}$ and $\Omega_{0}$, of the gate operation pulses are shown in Figure 2(a), 2(b), 2(c), and 2(d) for $\sigma_x$, $\sigma_y$, $\sigma_z$, and Hadamard gates, respectively. Pulse envelopes for $\sigma_x$ and $\sigma_y$ gates shown in Figure 2(a) and Figure 2(b), are the same since $\theta$ values are the same. In addition, in both figures $\Omega_1 = -\Omega_0$ as $B = -A$. For $\sigma_z$ gate, $\Omega_0 = 0$, which comes from the fact that $\theta = 0$. $\Omega_1$ in Figure 2(c) differs from that in Figure 2(d) by about 8\% which is due to different values of $\theta$ as shown in Table I. In all cases the instantaneous Rabi frequency does not exceed 1.5 MHz, which is a realistic value in experiments. The phase compensation pulses have the same envelopes except for exchanging the two fields and amplifying their magnitude by a factor of 2, respectively.

\begin{table} [htbp] 
  \centering
  \caption{Optimized $a_n$ parameters for all gate operations.}
  \label{Table 3}}
  \centering
  \setlength{\tabcolsep}{4mm}{
  \scalebox{0.8}{
  \begin{tabular}{c c c c c c c c c} 
  \hline
  \hline
   & $a_1$ & $a_2$ & $a_3$ & $a_4$ & $a_5$ & $a_6$ & $a_7$ & $a_8$    \\
  \hline
  Op1 & 0.0246 & -0.8980 & 0.0066 & 0.3668 & -0.0021 & -0.1358 & -0.0048 & 0.0179 \\
  \hline
  Op2 & -0.5400 & -0.1582 & 5.7637  &  3.9338 &  -0.6641 &  -0.6328 & -1.9186 & -1.5777 \\
  \hline
  \hline
  \end{tabular} }
\end{table}

\begin{figure}[H]
\begin{minipage}{8cm}
\centering
	\includegraphics[width=6.5cm,height=4.5cm]{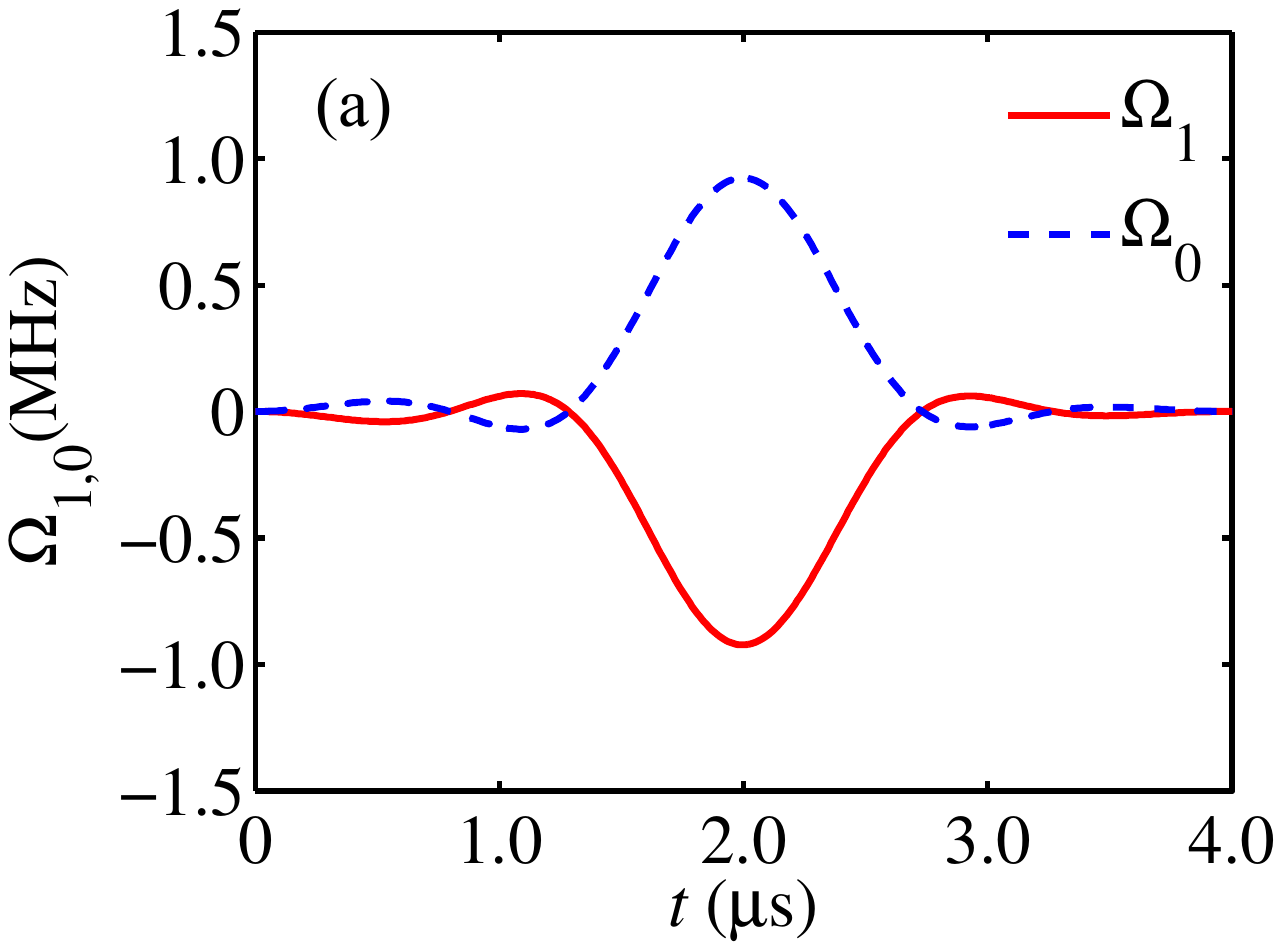} %
\end{minipage}
\hfill  
\begin{minipage}{8cm}
\centering
	\includegraphics[width=6.5cm,height=4.5cm]{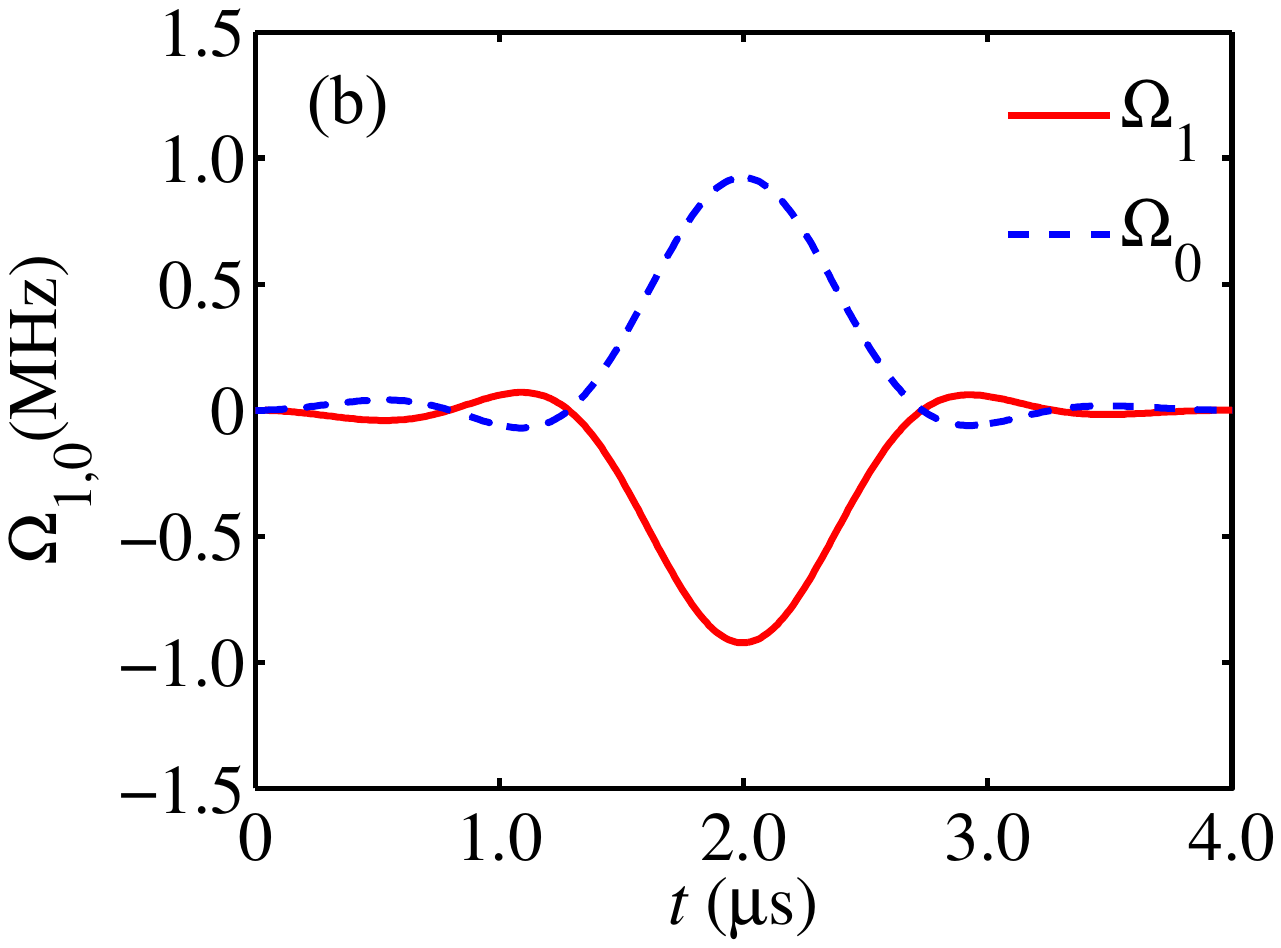}
\end{minipage}
\vfill
\begin{minipage}{8cm}
\centering
	\includegraphics[width=6.5cm,height=4.5cm]{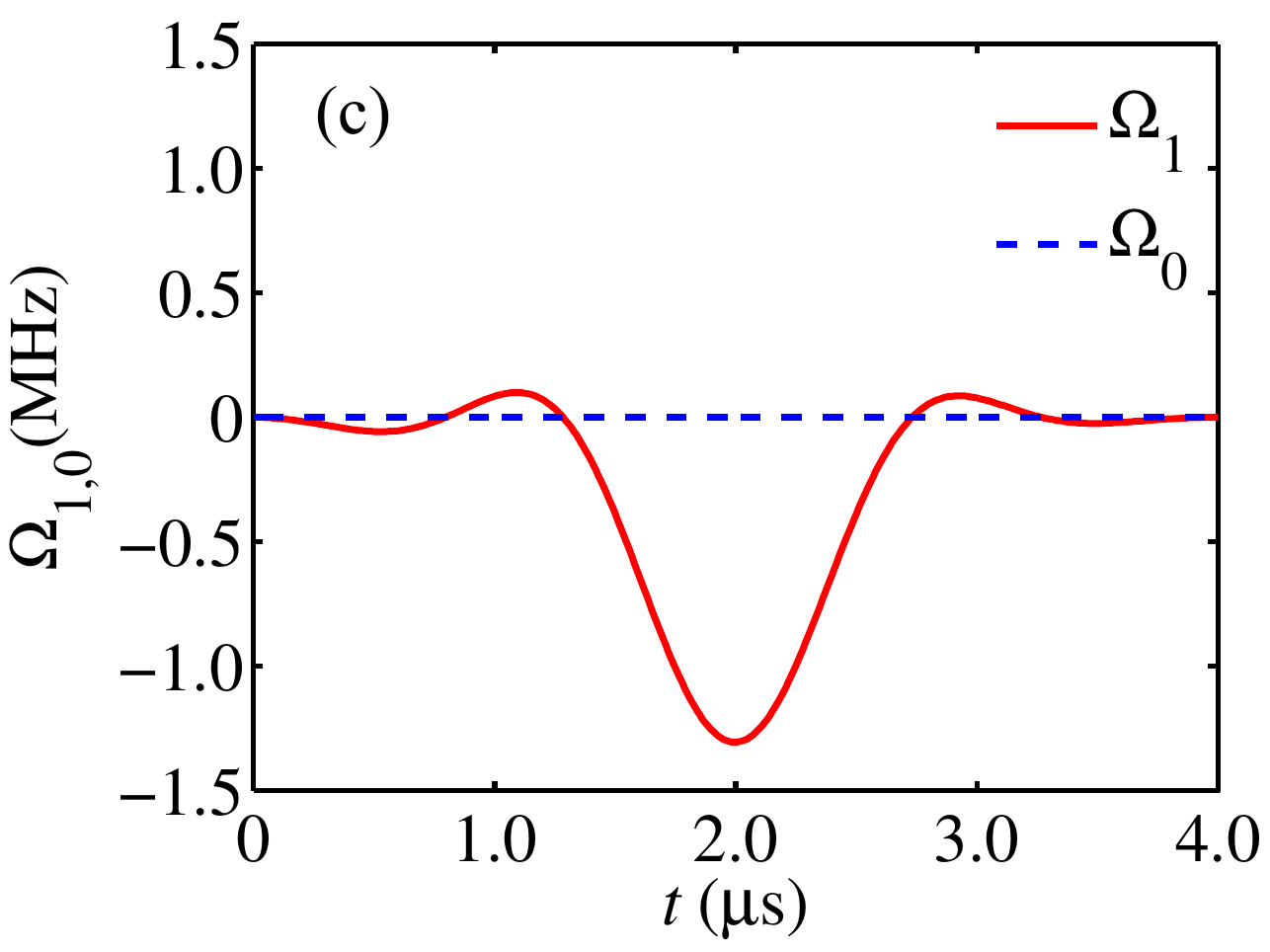}
\end{minipage}
\hfill
\begin{minipage}{8cm}
\centering
	\includegraphics[width=6.5cm,height=4.5cm]{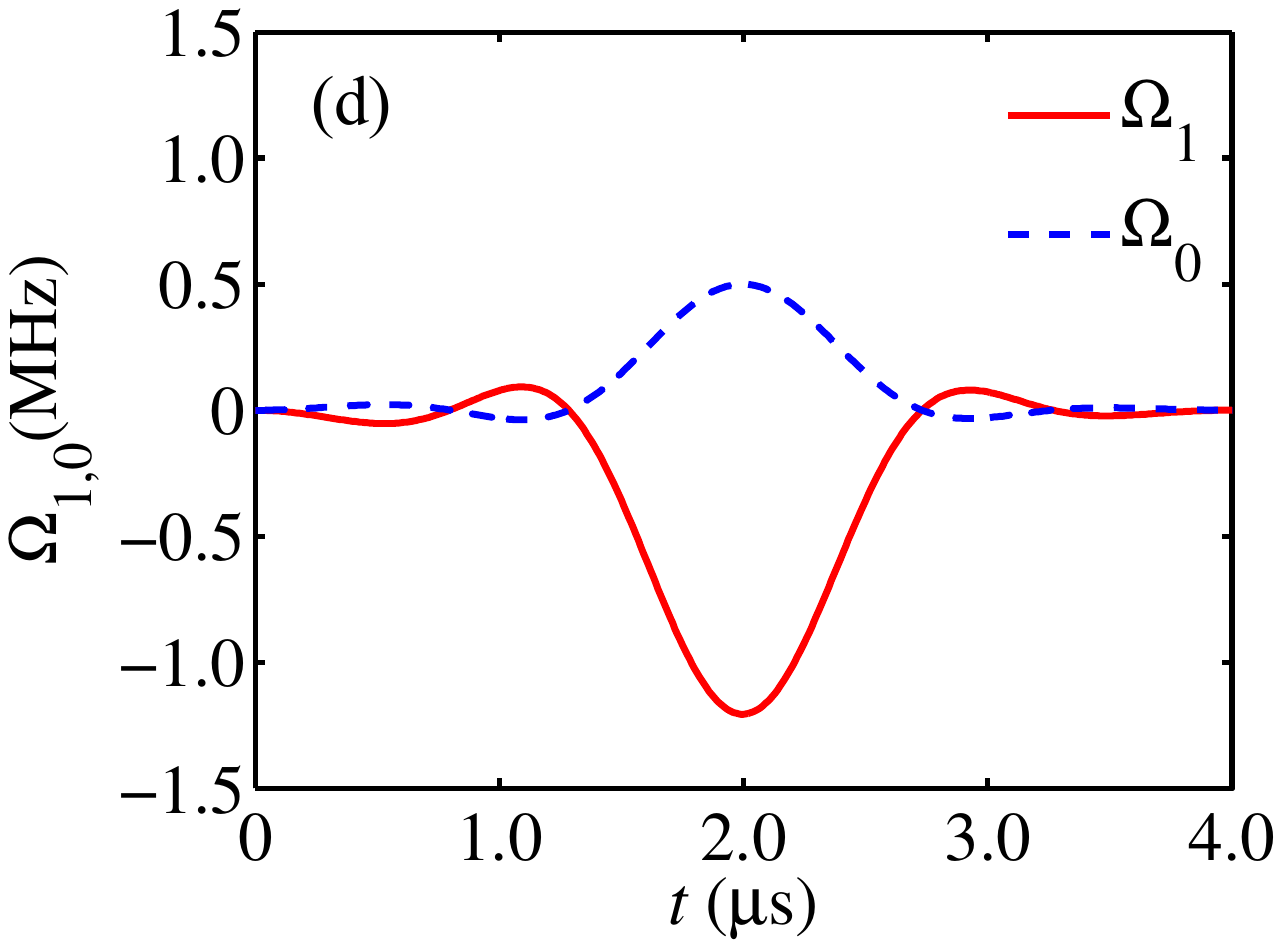}
\end{minipage}
	\caption{(Color online) Envelope of Rabi frequencies of pulses for (a) $\sigma_x$ gate, (b) $\sigma_y$ gate, (c) $\sigma_z$ gate, and (d) Hadamard gate. $\Omega_{1,0}$ denotes the Rabi frequency for transition $\left|1\right>\rightarrow\left|\rm{e}\right>$ and $\left|0\right>\rightarrow\left|\rm{e}\right>$.}
	\label{fig.2}
\end{figure} 

The time evolution of the quantum state in response to the above pulses are shown in Figure 3. Population starts from $\left|1\right>$ state, and ends at $\left|0\right>$ for $\sigma_x$ and $\sigma_y$ gate, at $\left|1\right>$ for $\sigma_z$, and half population on both $\left|0\right>$ and $\left|1\right>$ for Hadamard gate, as expected.

\begin{figure}[H]
\begin{minipage}{8cm}
\centering
	\includegraphics[width=6.5cm,height=4.5cm]{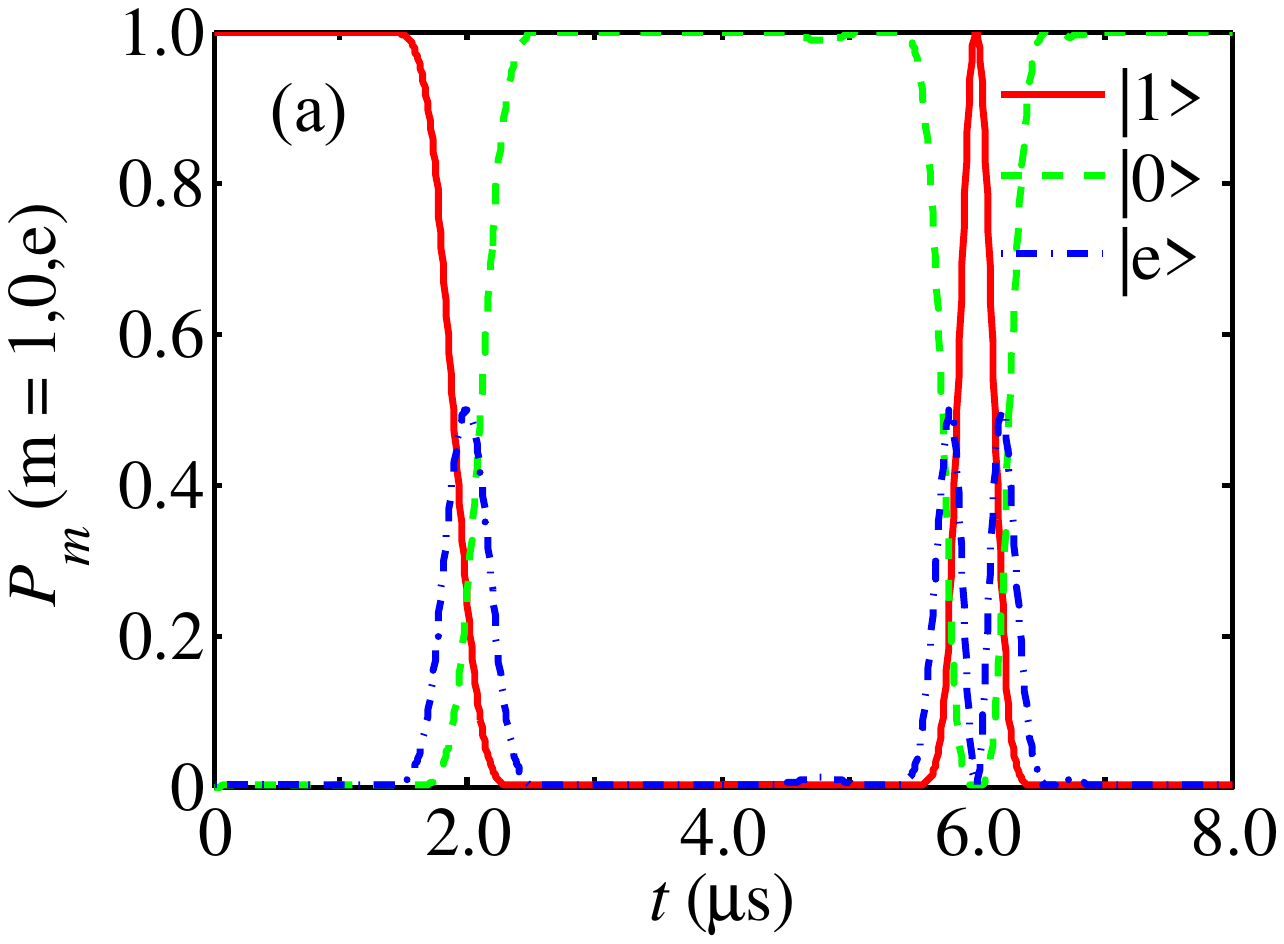}
\end{minipage}
\hfill  
\begin{minipage}{8cm}
\centering
	\includegraphics[width=6.5cm,height=4.5cm]{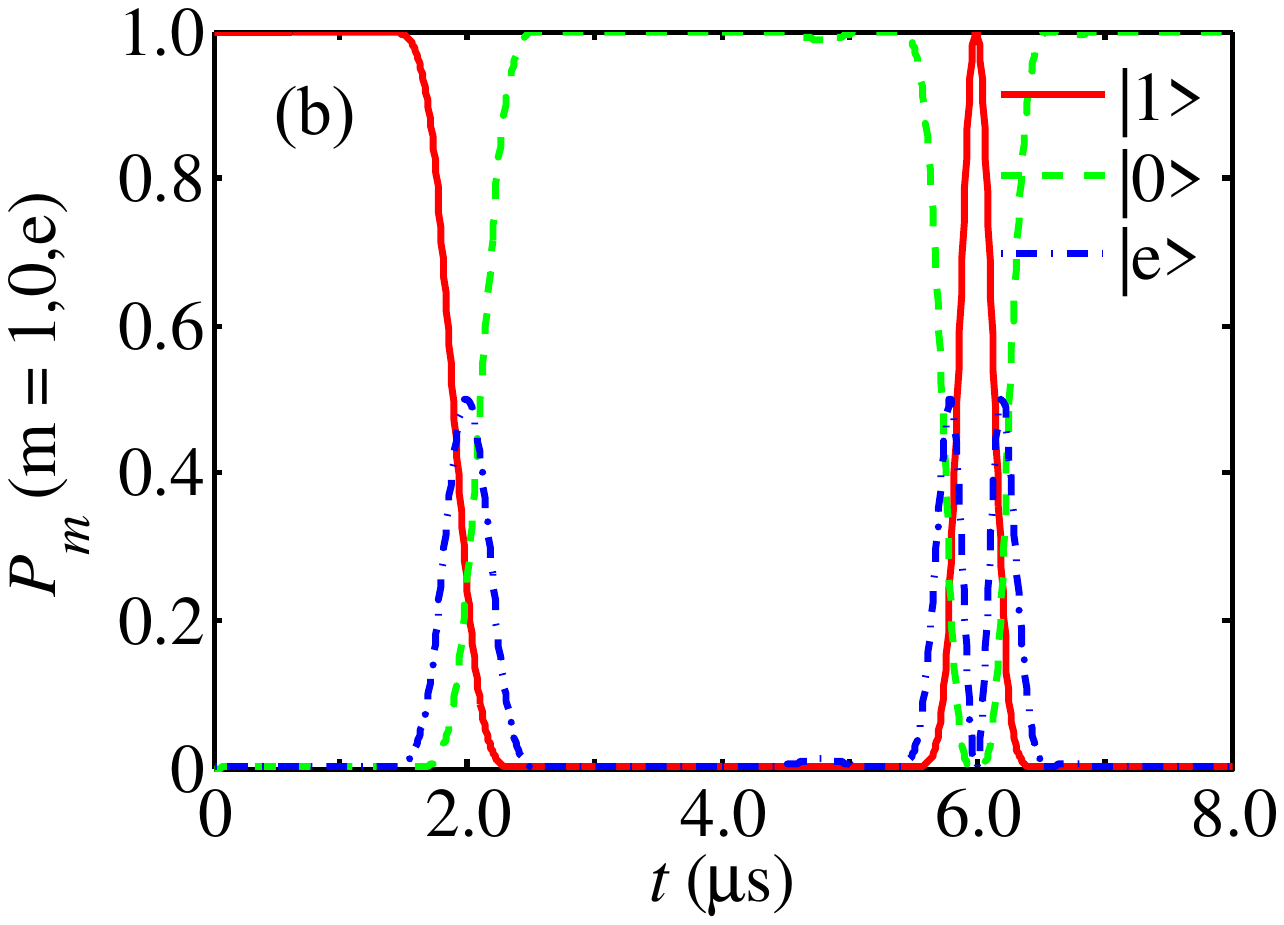}
\end{minipage}
\vfill
\begin{minipage}{8cm}
\centering
	\includegraphics[width=6.5cm,height=4.5cm]{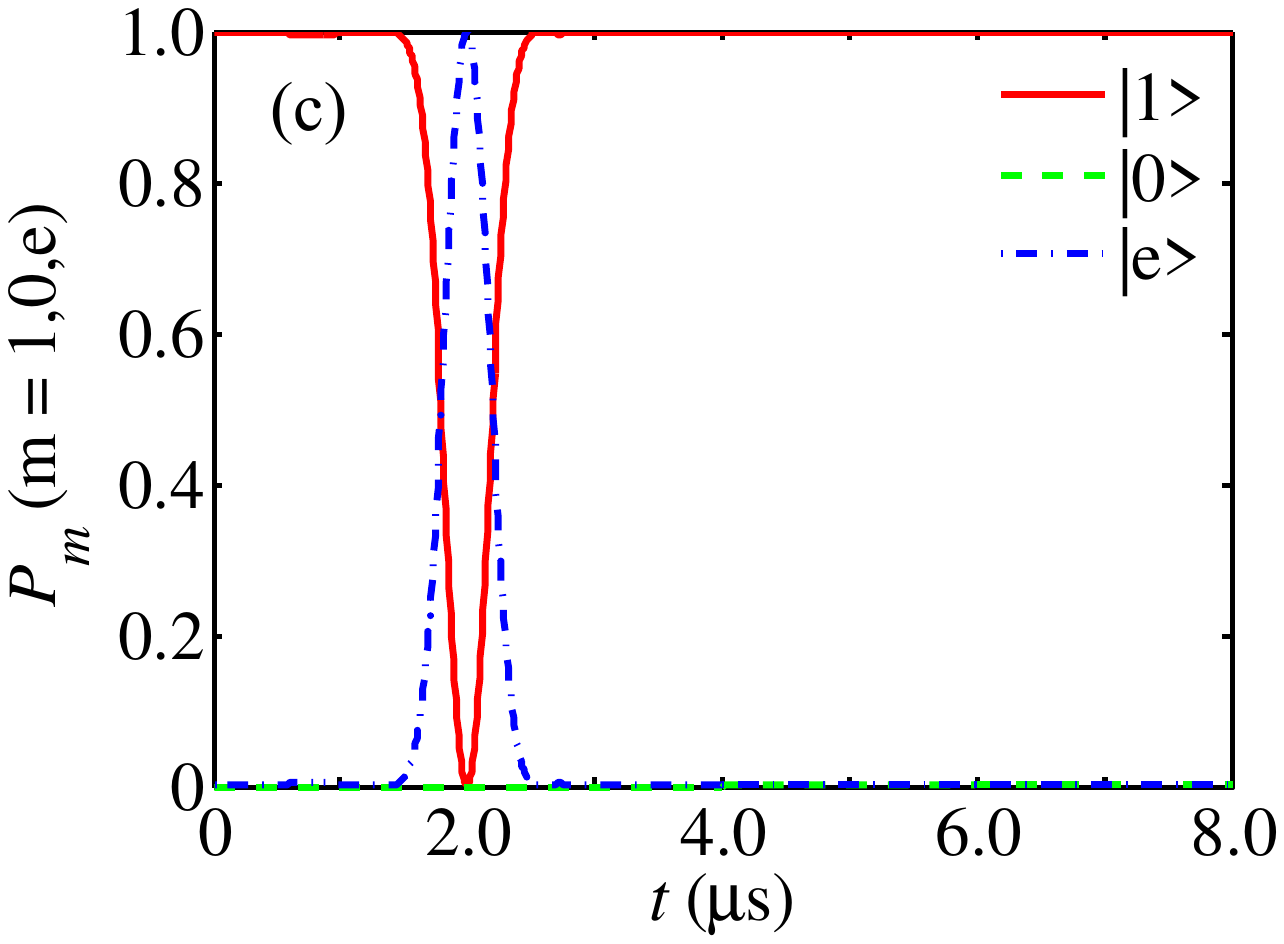}
\end{minipage}
\hfill
\begin{minipage}{8cm}
\centering
	\includegraphics[width=6.5cm,height=4.5cm]{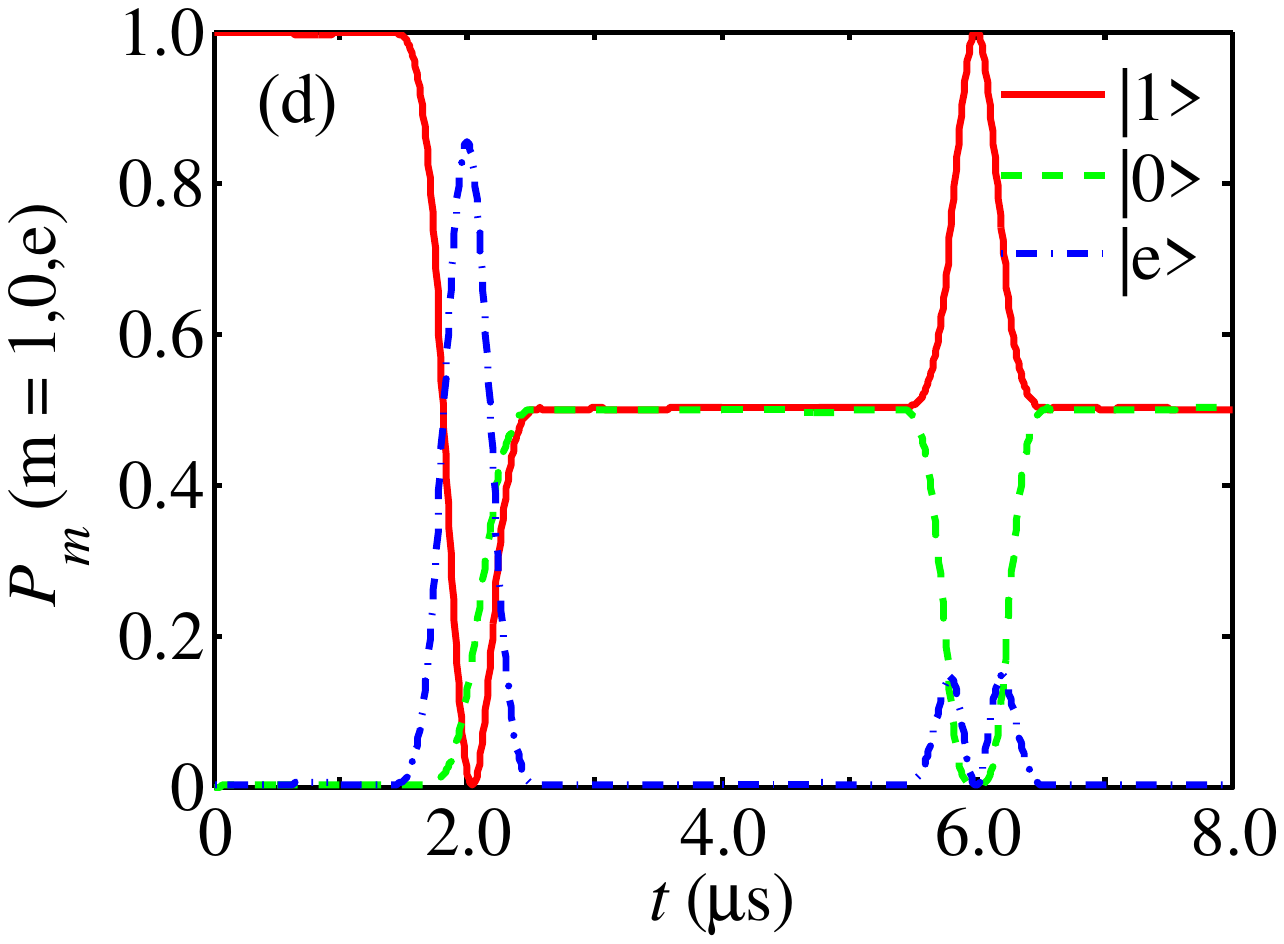}
\end{minipage}
	\caption{(Color online) State evolutions of (a) $\sigma_x$ gate, (b) $\sigma_y$ gate, (c) $\sigma_z$ gate, and (d) Hadamard gate.}
	\label{fig.3}
\end{figure} 

The operational fidelity $F$, as described by Eq. (7), of each gate operation behaves differently at different frequency detuning $\Delta$. The dependence is shown as the solid-red curve in Figure 4. An average fidelity of more than 99\% is accomplished within a detuning range as large as $\pm$410 kHz in all cases. For comparison, the fidelity obtained with Gaussian pulses with the same duration and the same full width at half maximum of Rabi frequencies in time ($t_{fwhm}$) are shown as dashed-blue curves; the dash-dotted-green curves show the situation for square pulses with the same duration. Clearly, our pulses are more robust to frequency detuning than either Gaussian or square pulses. It is worth noting that our pulses' robustness can be further improved to $\pm$600 kHz at a price of higher instantaneous Rabi frequencies (12 MHz) if $a_n$ parameters listed as Op2 in Table III are used. The results are shown in Figure 5, where the average fidelity is 99.9\%. The reason why optimized $a_n$ parameters improve the robustness may be that for larger detuning frequencies, the polarization induced by the pulse changes linearly with the amplitude of the Fourier transform of the pulses at that frequency. However, for small detuning frequencies, this change is nonlinear. This nonlinear behaviour may be compensated for by the optimal $a_n$ values, thus providing a nearly flat response across the zero detuning point \cite{Warren1984}.

\begin{figure}[H]
\begin{minipage}{8cm}
\centering
	\includegraphics[width=6.5cm,height=4.5cm]{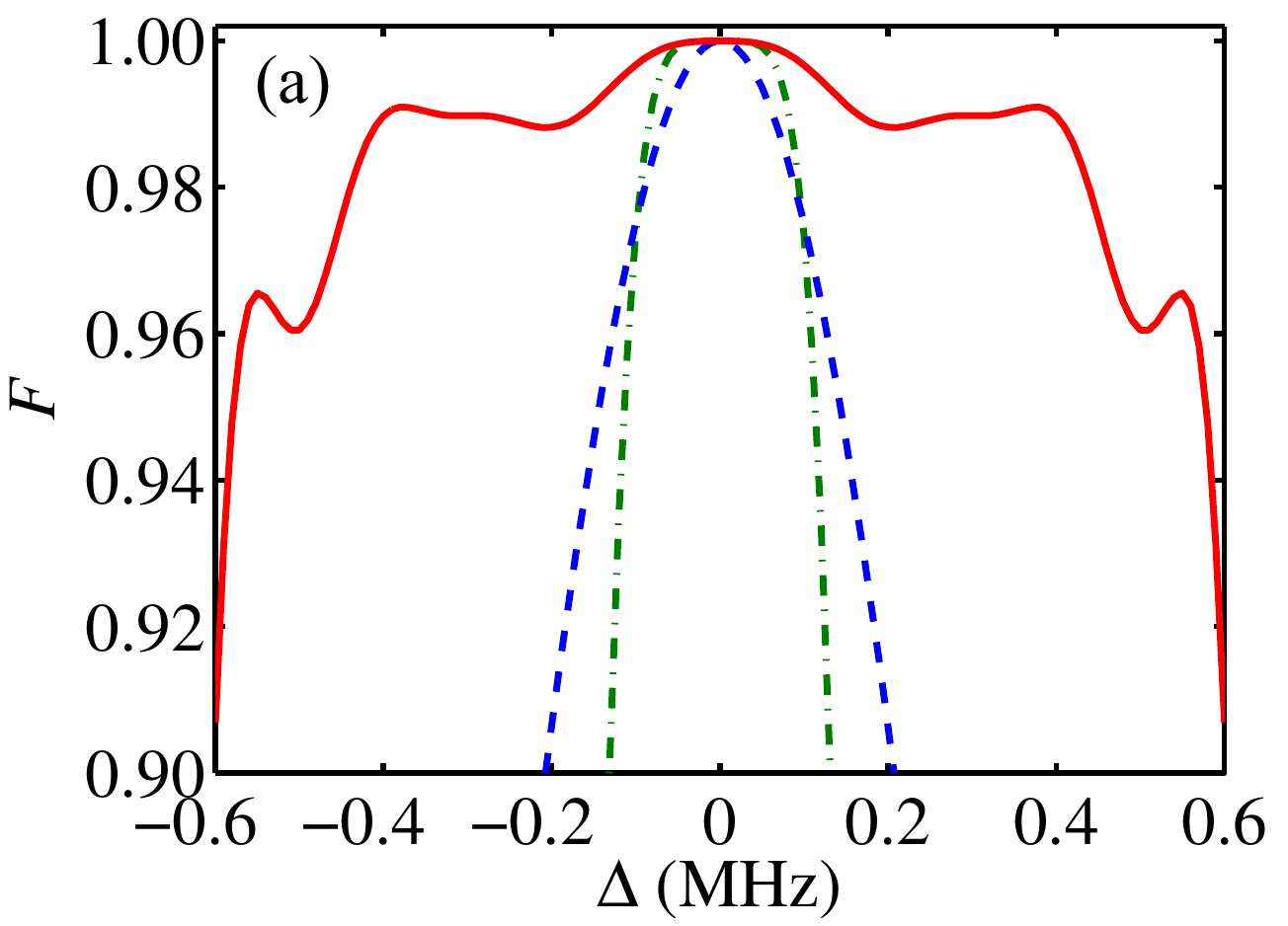}
\end{minipage}
\hfill  
\begin{minipage}{8cm}
\centering
	\includegraphics[width=6.5cm,height=4.5cm]{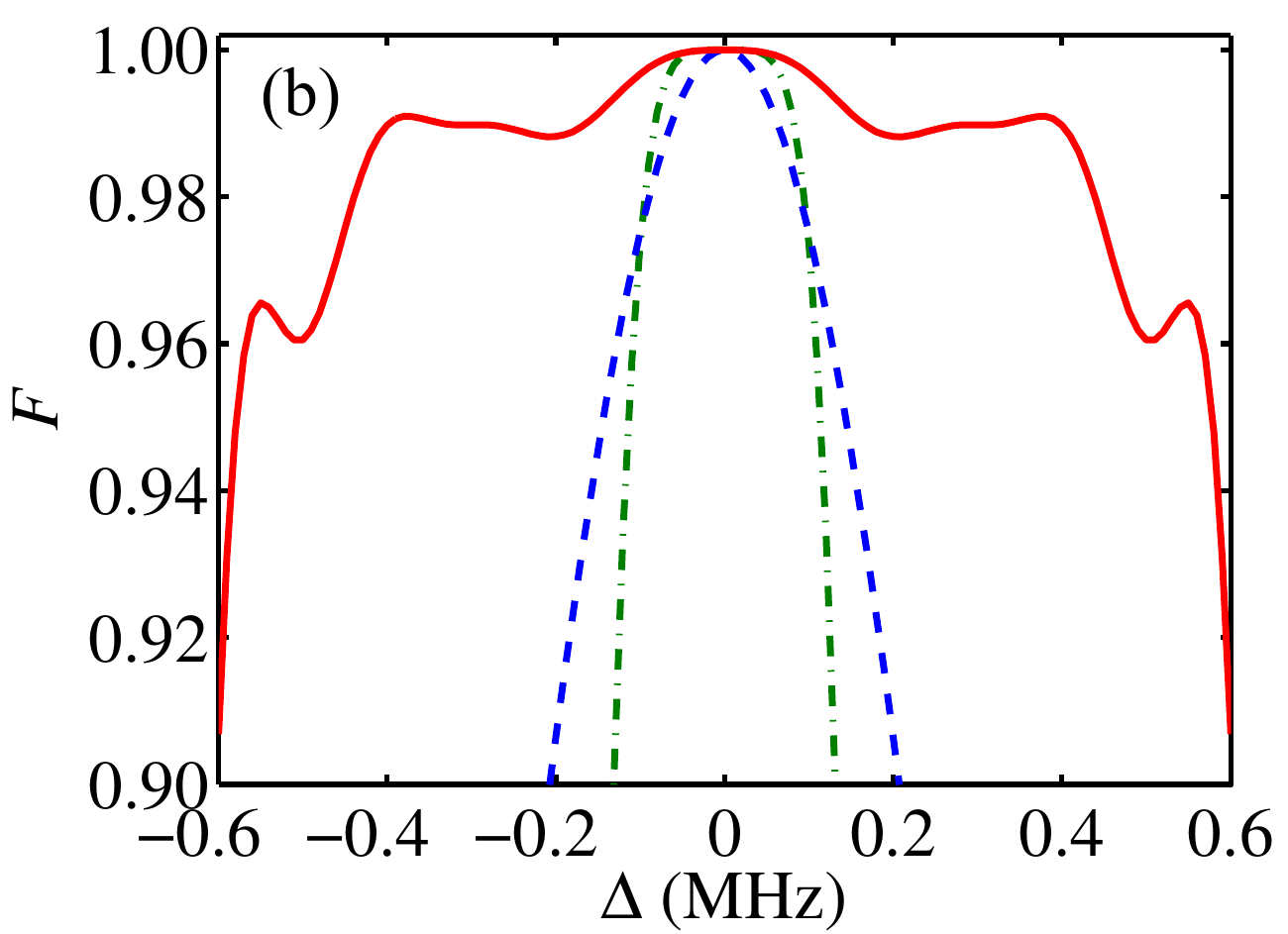}
\end{minipage}
\vfill
\begin{minipage}{8cm}
\centering
	\includegraphics[width=6.4cm,height=4.5cm]{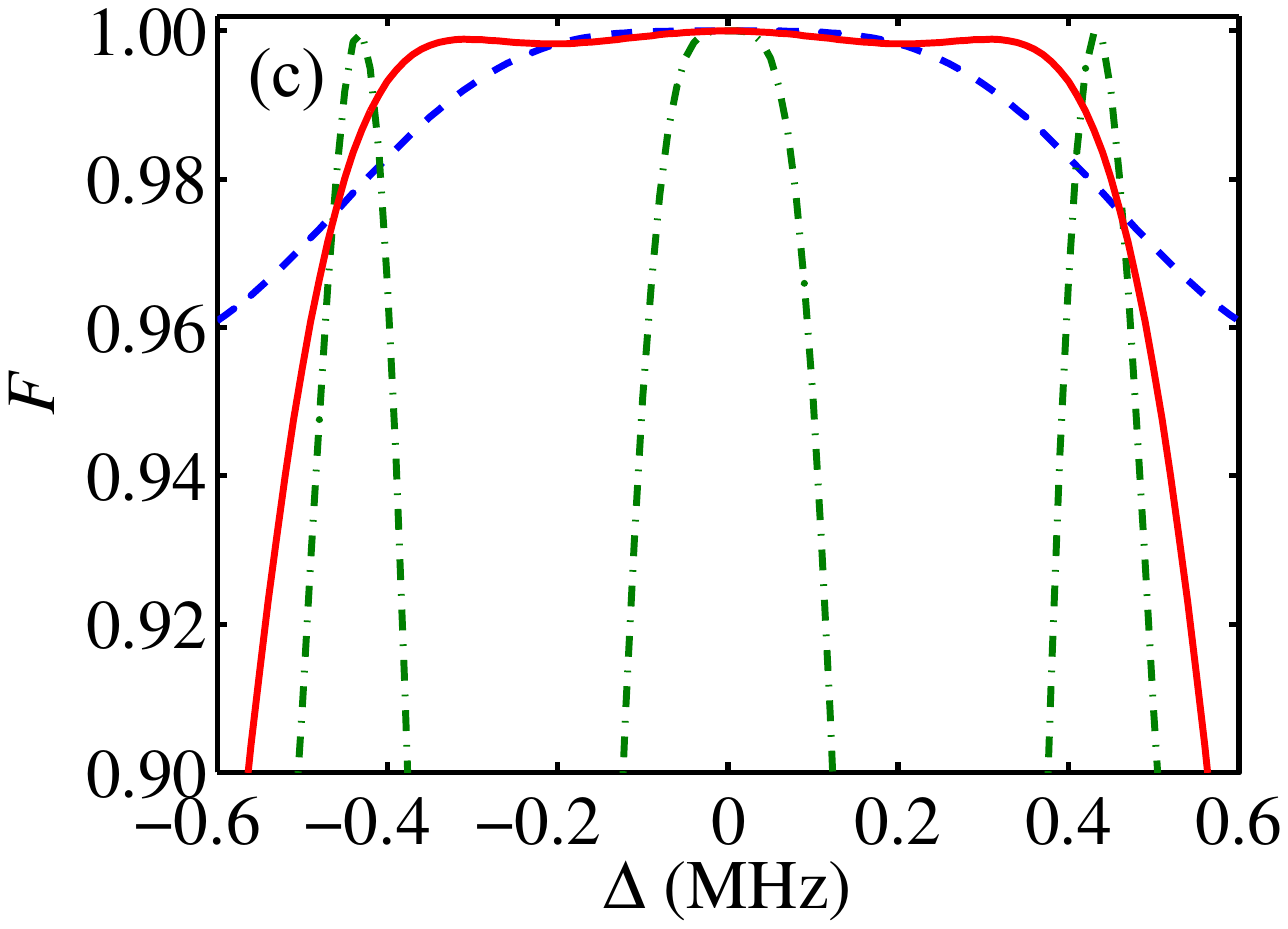}
\end{minipage}
\hfill
\begin{minipage}{8cm}
\centering
	\includegraphics[width=6.5cm,height=4.5cm]{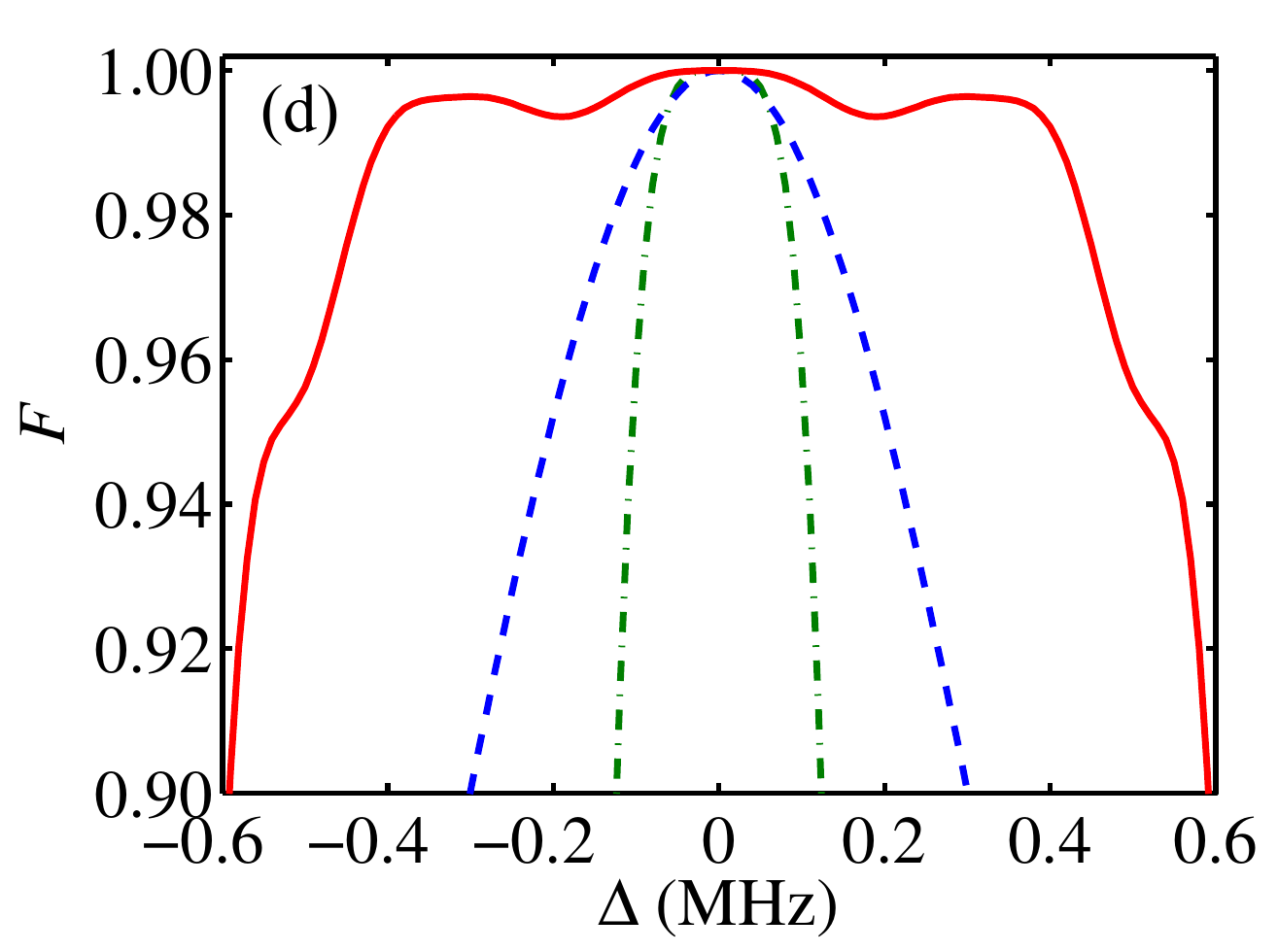}
\end{minipage}
	\caption{(Color online) Dependence of fidelity on frequency detuning for (a) $\sigma_x$ gate, (b) $\sigma_y$ gate, (c) $\sigma_z$ gate, and (d) Hadamard gate. Red-solid curves: fidelity obtained with the pulses described in this work with optimized $a_n$ parameters shown as Op1 in Table III. Dashed-blue curves: performance of Gaussian pulses with the same $t_{fwhm}$ and pulse duration as our pulses. Dash-dotted-green curves: performance of square pulses with the same duration.}
	\label{fig.4} 
\end{figure} 

\begin{figure}[H]
\begin{minipage}{8cm}
\centering
	\includegraphics[width=6.5cm,height=4.5cm]{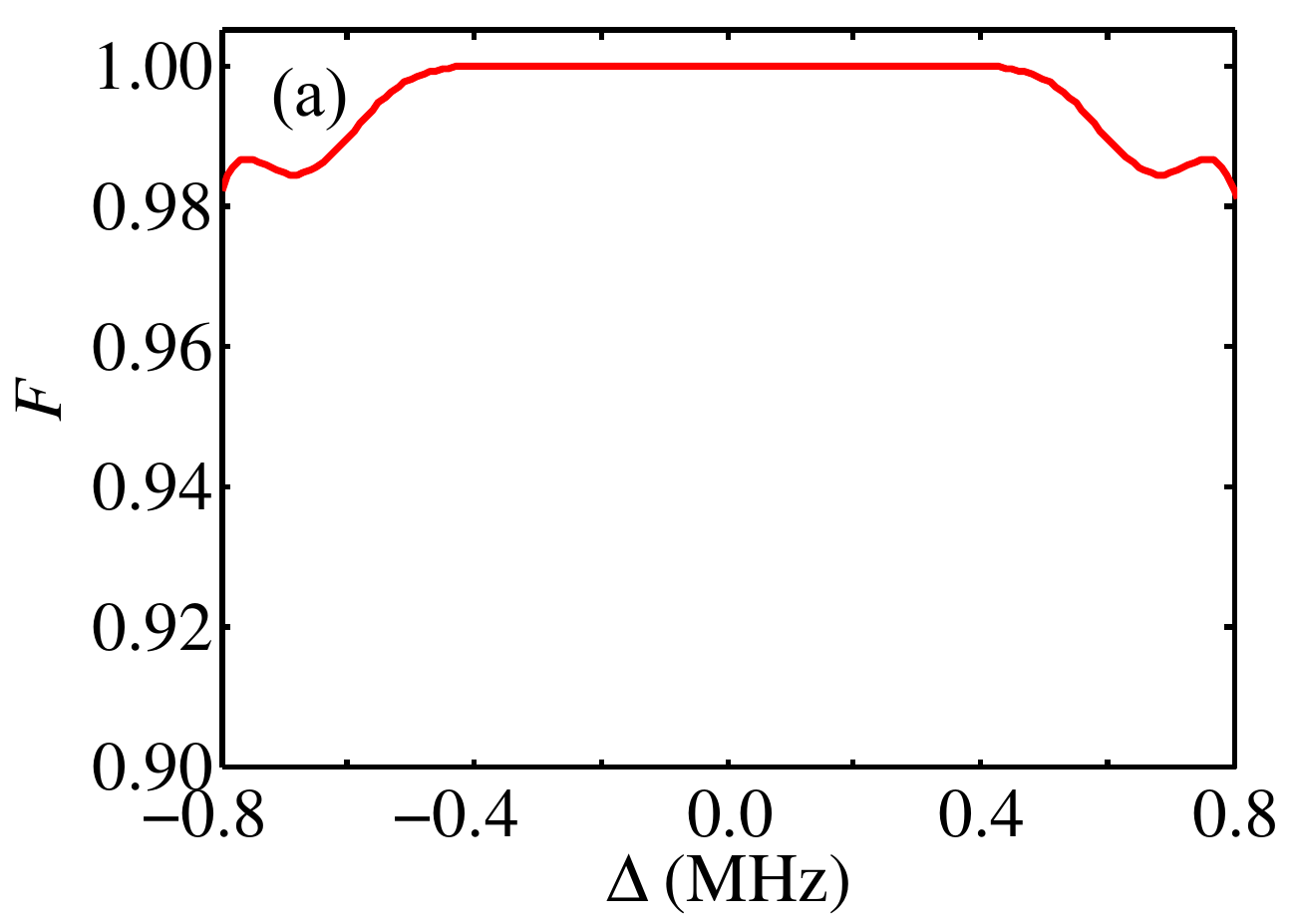}
\end{minipage}
\hfill  
\begin{minipage}{8cm}
\centering
	\includegraphics[width=6.5cm,height=4.5cm]{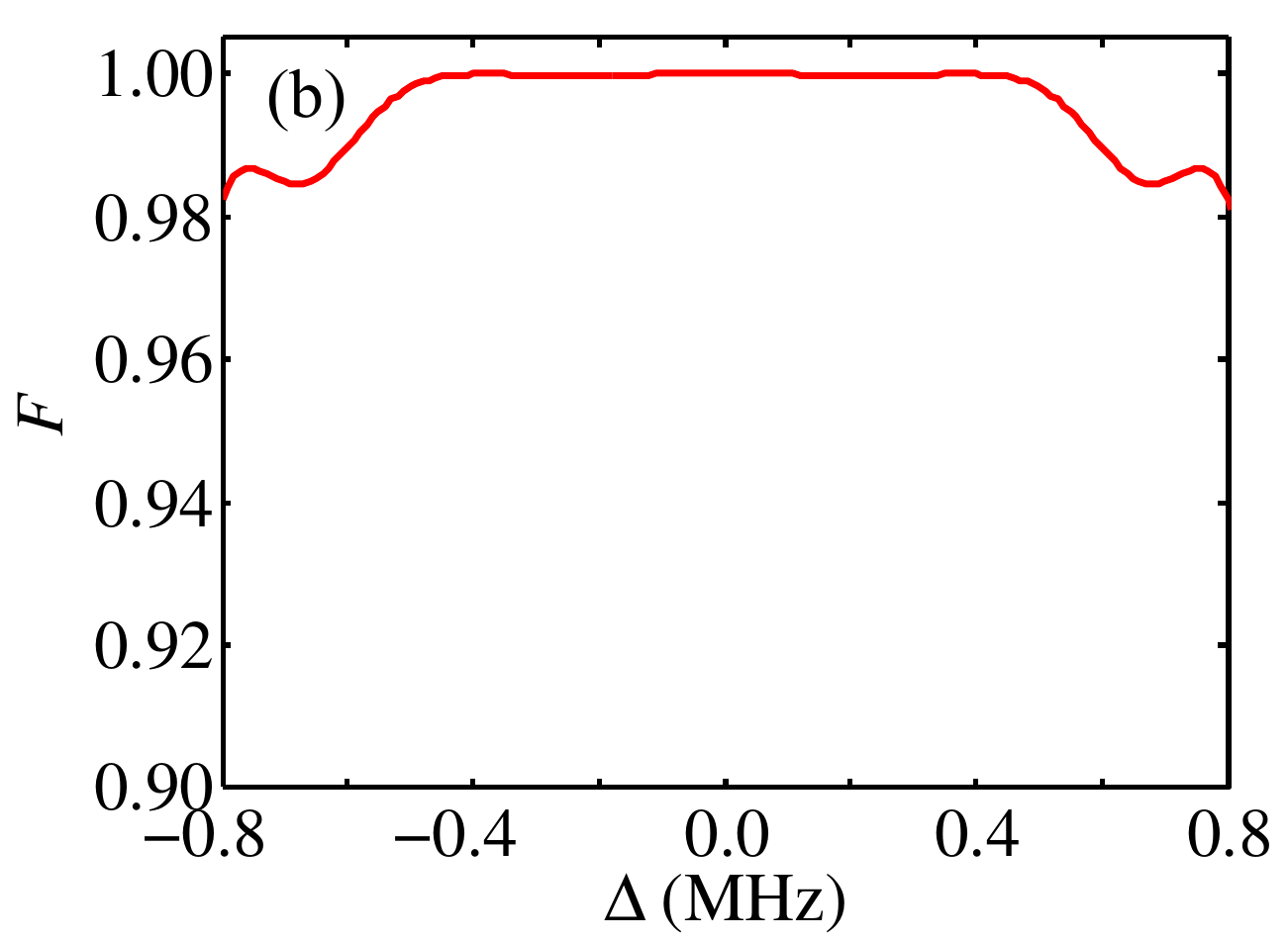}
\end{minipage}
\vfill
\begin{minipage}{8cm}
\centering
	\includegraphics[width=6.5cm,height=4.5cm]{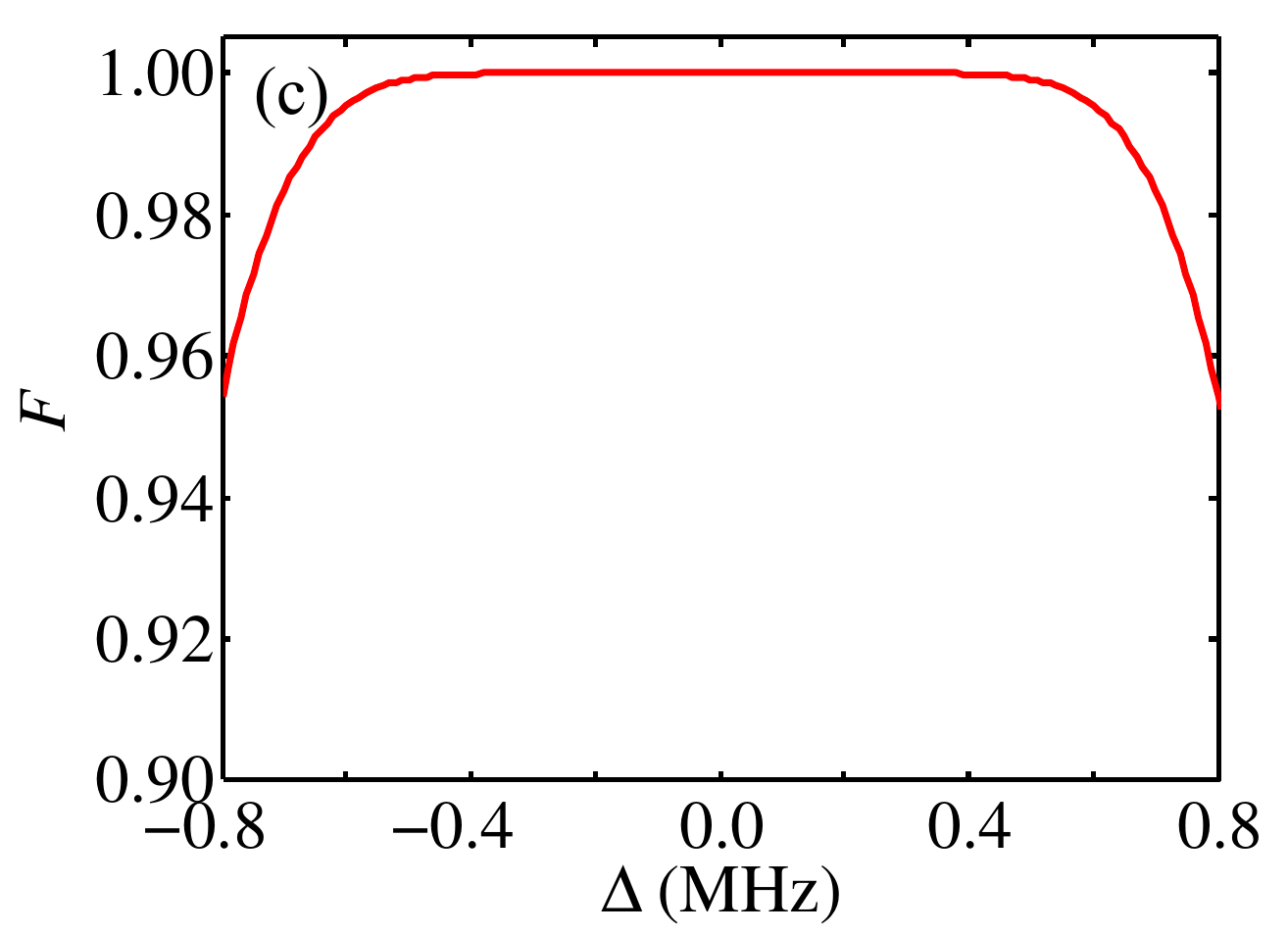}
\end{minipage}
\hfill
\begin{minipage}{8cm}
\centering
	\includegraphics[width=6.5cm,height=4.5cm]{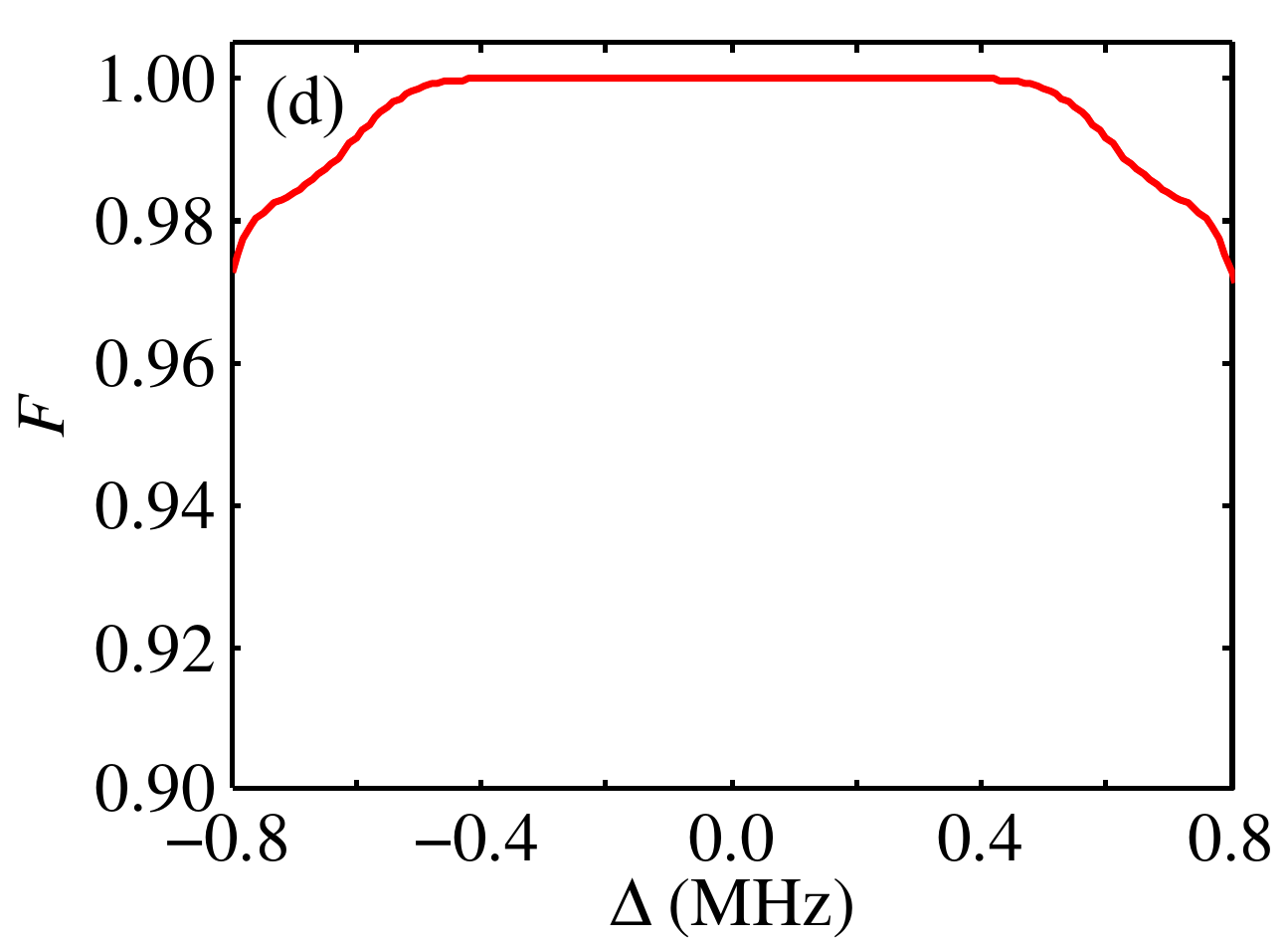}
\end{minipage}
	\caption{(Color online) Dependence of fidelity on frequency detuning using $a_n$ parameters shown as Op2 in Table III for (a) $\sigma_x$ gate, (b) $\sigma_y$ gate, (c) $\sigma_z$ gate, and (d) Hadamard gate. }
	\label{fig.5}
\end{figure} 

Either microwave or light pulses shown in Figure 2 can be used for gate operations in a quantum system. Microwave pulses can be directly generated with an arbitrary waveform generator (AWG). Light pulses can be created from the 1st order deflected beam of an acousto-optical modulator (AOM), the typical rise time of which is tens of nanoseconds for a beam focal diameter of 100 $\rm{\mu}$m. The AOM is driven by two radio frequency (RF) signals that can be generated with an AWG. The amplitude ($E_{1,0}$), frequencies ($f_{1,0}$), and phase ($\varphi_{1,0}$) of the two RF signals should satisfy: 
\begin{equation}
\begin{aligned}
&E_{1,0}  =  C \cdot \left\{
\begin{array} {rcl}
&\Omega_{1,0}, & \Omega_{1,0} > 0\\
&e^{i\pi}\cdot|\Omega_{1,0}|, & \Omega_{1,0} < 0\\
\end{array}
\right.,\\
&f_1 - f_0 = f_{10},\quad \rm{and}\\
&\varphi_1 = 0, \quad\varphi_0 = -\varphi,
\end{aligned}
\end{equation} where $C$ represents a constant accounting for the conversion factor from the amplitude of electric field of a RF signal to that of the light pulse, and $f_{10}$ is the frequency interval between two qubit energy levels. Depending on both the time and amplitude resolution, and on the noise characteristics of the AWG, the generated waveform may deviate from the ideal one that is constructed from Eq. (2) using $a_n$ values shown in Table III. Therefore it is interesting to investigate how the operational fidelity changes in response to variations in $a_n$ around their optimal values. However, from Eq. (2) it is not straightforward to identify which variation of $a_n$ yields the largest deviation in the waveform. As a qualitative illustration, here we consider a simple yet instructive case where a variation occurs only on $a_2$ while other $a_n$ are fixed, i.e. $\Delta a_2\neq0$ and $\Delta a_n = 0$ ($n\neq2$). The reason for this is that, among all optimal $a_n$ parameters, $a_2$ has the largest weight so its variation strongly affects the Rabi frequency and the fidelity. The results are shown in Figure 6, where $\eta = \Delta a_2/a_2$ denotes the fractional variation. In all cases, fidelity at $\Delta = 0$ is not affected by the variation. This is because the pulse area of $\Omega(t)$ is independent of $a_n$, and at $\Delta = 0$, $\left|\psi(t_2)\right> = U^{(1)}(t_1,0)\left|\psi(0)\right> \equiv \left|\psi_{\rm{tg}}\right>$ holds. For $\Delta\in[-60, 60]$ kHz, $F$ is robust against variations within $\pm$30\%. Robustness decreases significantly at larger detuning frequencies, which means that an AWG that can provide more accurate control over $a_n$ is preferable. 

\begin{figure}[H]
\begin{minipage}{8cm}
\centering
	\includegraphics[width=6.5cm,height=4.5cm]{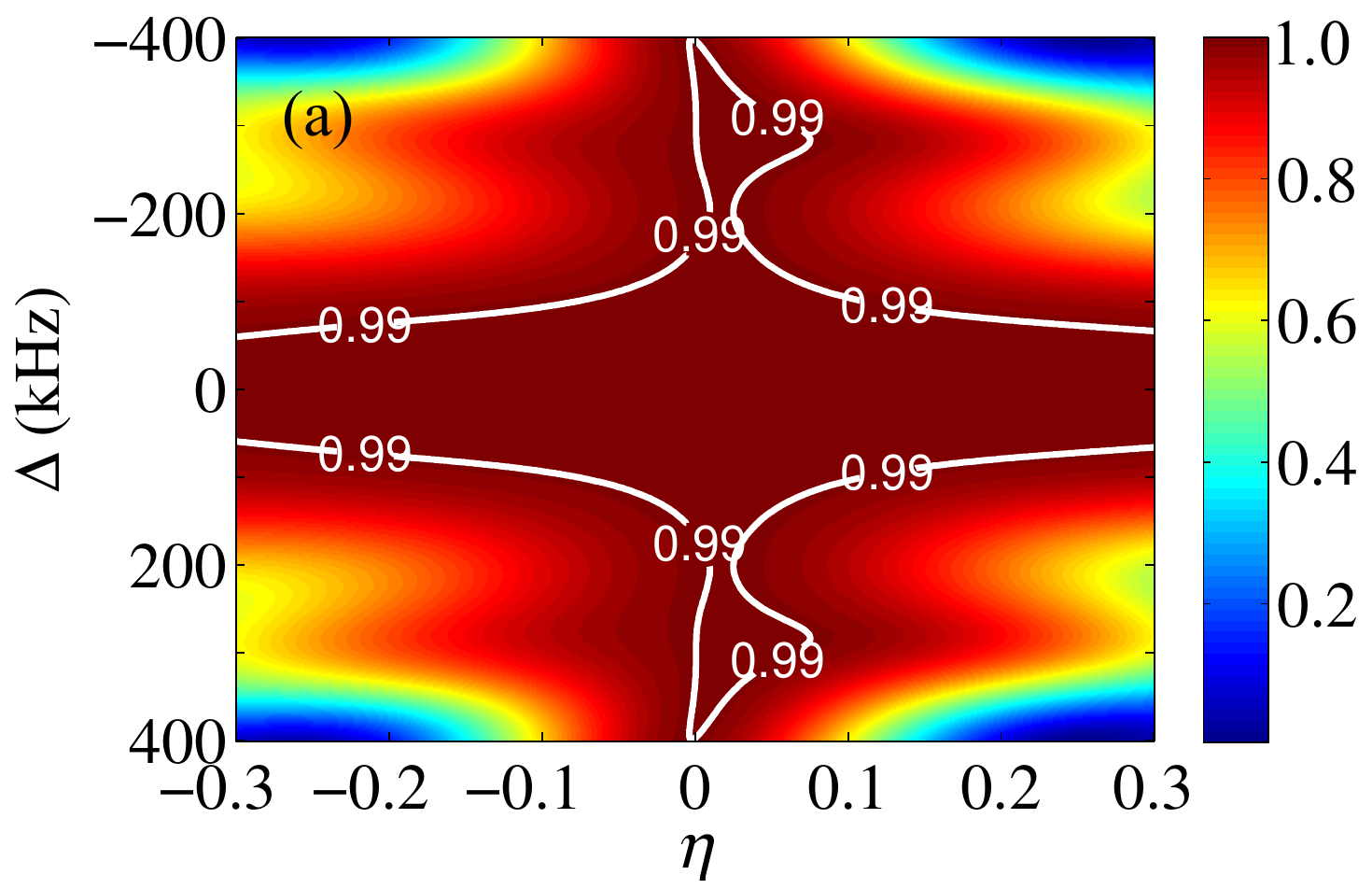}
\end{minipage}
\hfill  
\begin{minipage}{8cm}
\centering
	\includegraphics[width=6.5cm,height=4.5cm]{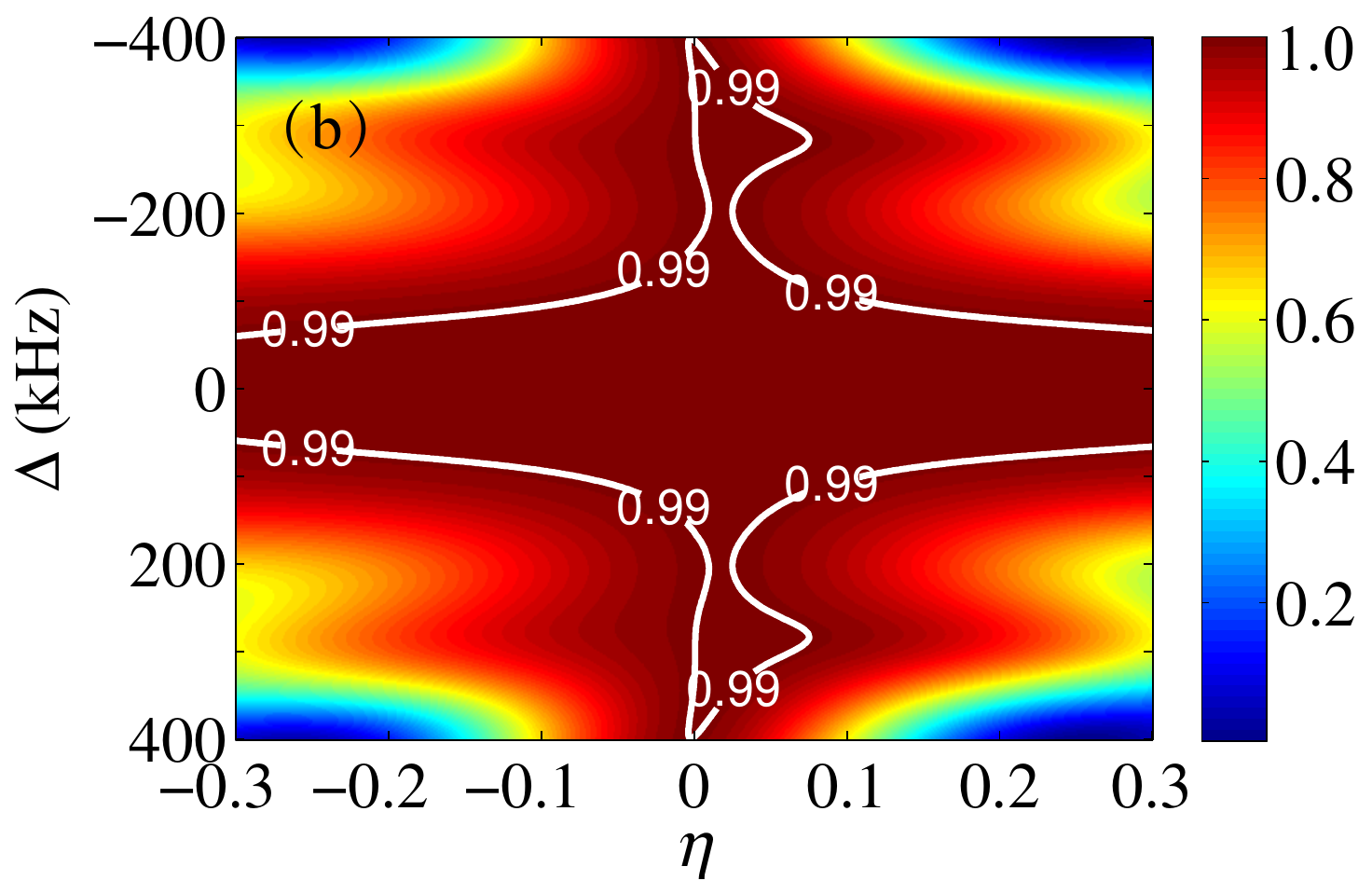}
\end{minipage}
\vfill
\begin{minipage}{8cm}
\centering
	\includegraphics[width=6.5cm,height=4.5cm]{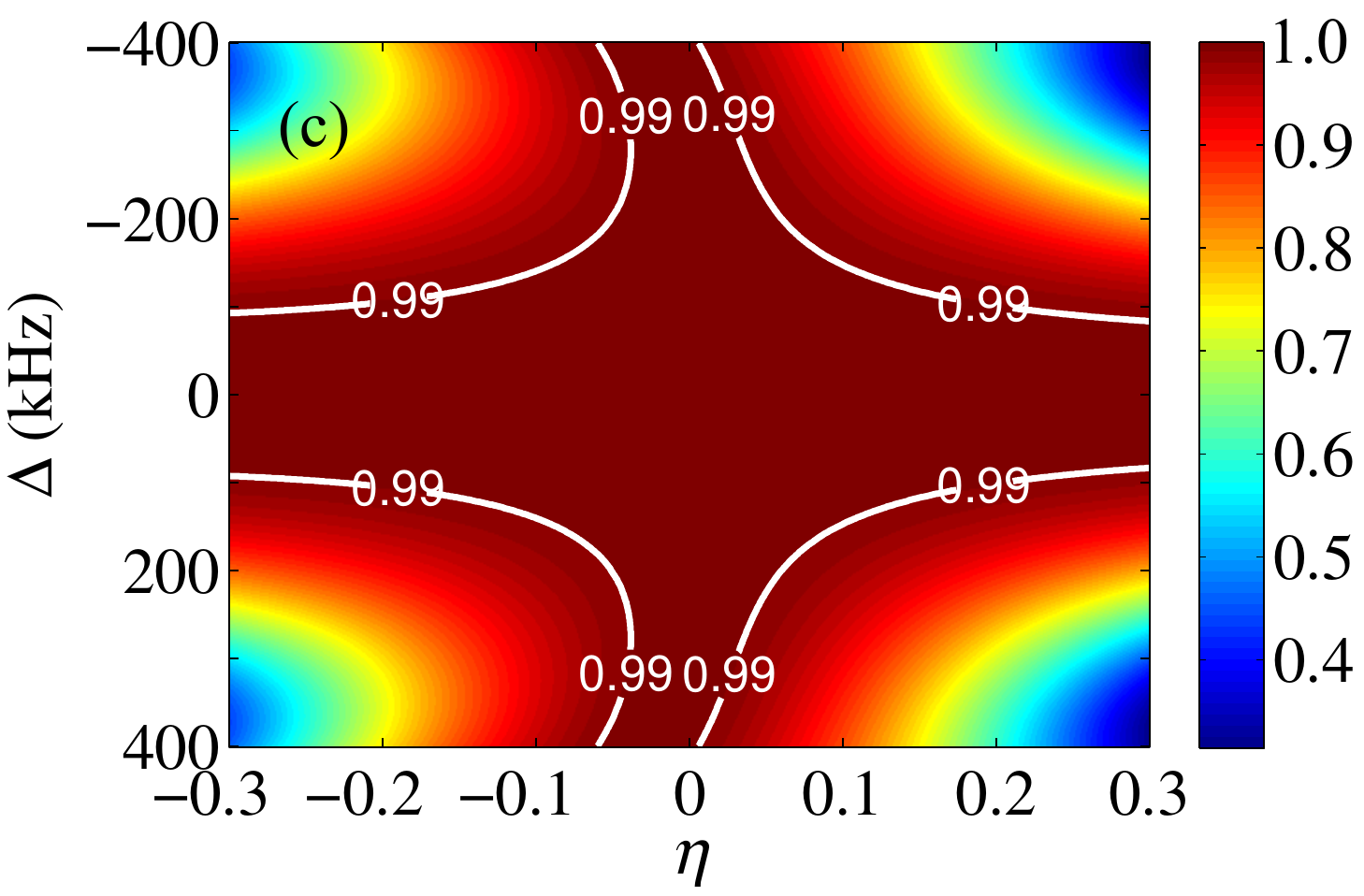}
\end{minipage}
\hfill
\begin{minipage}{8cm}
\centering
	\includegraphics[width=6.5cm,height=4.5cm]{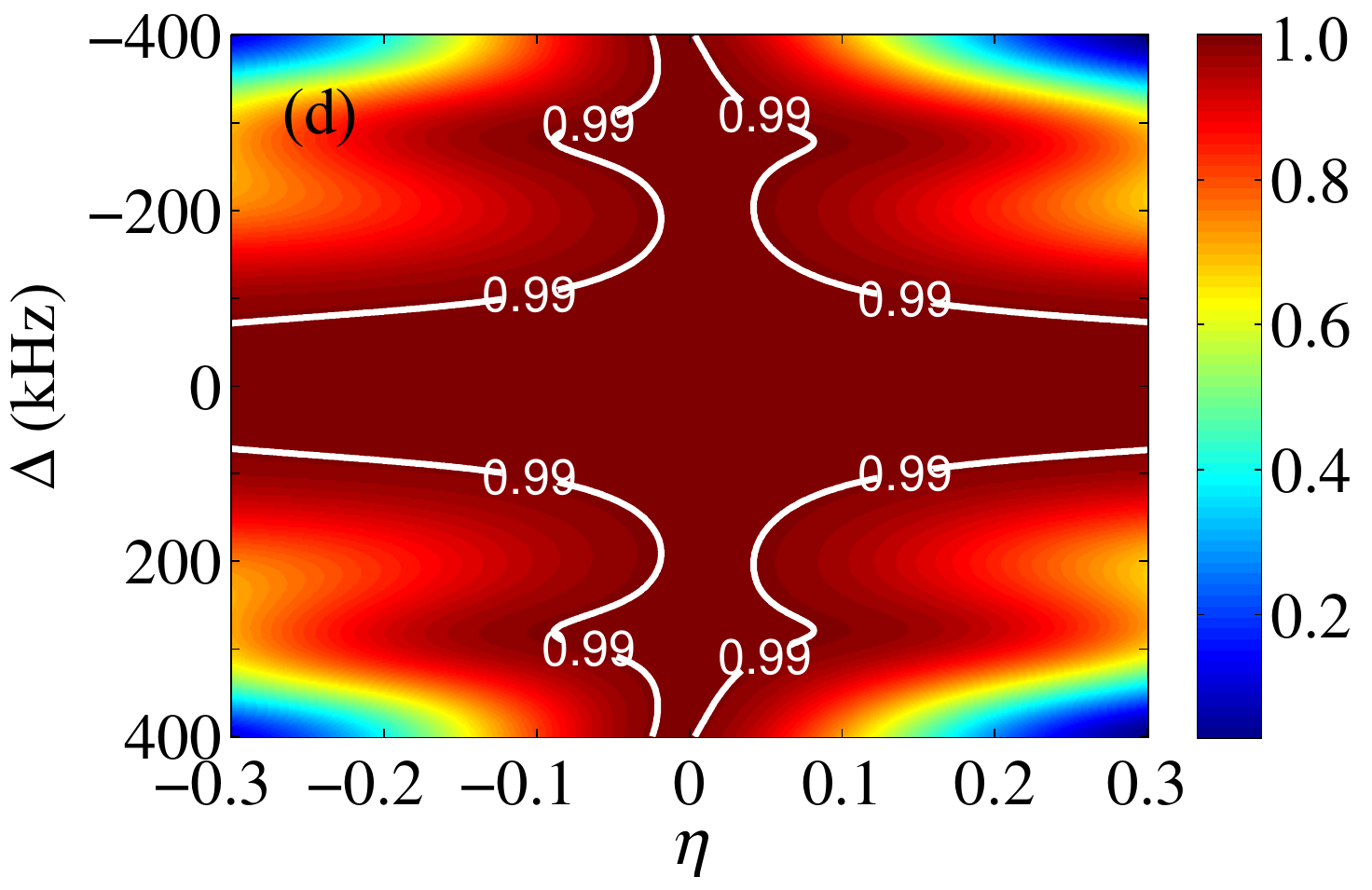}
\end{minipage}
	\caption{(Color online) Fidelity in response to variation on $a_2$ at $\Delta$ kHz frequency detuning for (a) $\sigma_x$ gate, (b) $\sigma_y$ gate, (c) $\sigma_z$ gate, and (d) Hadamard gate. White lines indicate those data points that $F = 0.99$.}
	\label{fig.6}
\end{figure} 

The method for constructing the robust gate-operation pulses presented in this work may be employed with the rare-earth-ion ensemble qubits, which have an inhomogeneous broadening of about 170 kHz \cite{Lars2008}. $a_n$ parameters would have to be optimized for this detuning range with the detuning distribution being considered. It may also be applied to the superconducting transmon qubits system for initializing the ancilla qubits. Their frquencies typically vary about 2\% \cite{Devoret2013, PC}, which corresponds to 100 MHz for a qubit transition frequency of 5 GHz. This robust range can be ensured if a pulse duration of $t_1 = 16$ ns is used with our pulses presented above.

\section{Discussion and Conclusion }

In this work we propose a method to design pulses based on a resonant system to achieve high fidelity gate operations in a slightly off-resonant system. The pulses are composed of a series of Cosine frequency components, the magnitude of which can be freely optimized to achieve robust fidelity. Our work shows that an average fidelity of 99$\%$ is achieved over a frequency detuning of $\pm$410 kHz for a pulse duration time of 4 $\mu$s. The frequency detuning range may extend to $\pm$600 kHz if higher Rabi frequencies can be employed. If higher Cosine harmonics can be used, the robust frequency range may even be increased. In addition, the robust frequency range can be adjusted by varying the pulse duration, as long as the Rabi frequencies are realistic.

We compare the robustness of the pulses shown in this work with that of Gaussian pulses, which have been used for geometric gate operations in experiments \cite{Abdumalikov2013}. For this, we use the same $t_{fwhm}$ and the same pulse duration. We also compare with square pulses having the same duration. Interestingly, our pulses show enhanced robustness compared to the other two. We investigate the dependence of operational fidelity on variations in $a_2$, and show that is robust against a 30\% variation for a frequency detuning range of [-60, 60] kHz. These pulses can be generated with an AWG, and may be used with both the rare-earth-ion ensemble qubits and superconducting transmon qubits system to significantly enhance the robustness of gate operations against the variations in frequency. Note that the optimized $a_n$ parameters might not be unique. In this work we only optimize them for the purpose of achieving robustness against frequency detuning. It is possible to optimize $a_n$ based on other problems, such as robustness against fluctuations in Rabi frequencies, minimization of Rabi frequencies, and insensitivity to spatial inhomogeneous distribution of intensity.

Our method to achieve the desired robustness via optimizing the pulse envelope provides an alternative method to design pulses for gate operations in a slightly off-resonant system. It may also effectively complement the perturbation theory that is often used to obtain an approximate solution when a physical quantity slightly deviates from a perfect situation.

\section*{ACKNOWLEDGEMENTS} 

The work was supported by National Natural Science Foundation of China (grant numbers 61505133 and 61674112), Natural Science Foundation of JiangSu Province (grant numbers BK20150308), Key Projects of Natural Science Research in JiangSu Universities (grant number 16KJA140001), the International Cooperation and Exchange of the National Natural Science Foundation of China NSFC-STINT (grant number 61811530020), the project of the Priority Academic Program Development (PAPD) of Jiangsu Higher Education Institutions, a JiangSu Province Professorship, and a Six Talent Peaks Project. Jie Lu acknowledges Shanghai Pujiang Program (grant numbers 16PJ1403100).

\bibliographystyle{prsty}

\end{document}